\documentclass[12pt]{article}
\input epsf 
\usepackage{epsfig,graphicx,amsmath}
\usepackage{epstopdf}
\usepackage{subcaption}
\usepackage[superscript]{cite}
\usepackage{cite}
\usepackage{enumerate}	
\usepackage[colorlinks=true, allcolors=blue]{hyperref}
\usepackage{multirow}
\usepackage{amssymb}
\usepackage{amsmath}
\numberwithin{equation}{section}

\graphicspath{{/Users/ninadvinayakmhatre/Documents/Desktop/Paper_4_Manuscript/pdf_images/}}
\usepackage{float}
\usepackage{color}
\usepackage{indentfirst}
\numberwithin{equation}{section}
\usepackage[utf8]{inputenc}
\usepackage[symbol]{footmisc}

\usepackage{titlesec}
\usepackage[title]{appendix}
\usepackage[section]{placeins}

\titleformat{\section}
  {\normalfont\fontsize{12}{12}\bfseries}{\thesection}{1em}{}
  
\titleformat{\subsection}
  {\normalfont\fontsize{12}{12}\filcenter\bfseries}{\thesubsection}{1em}{}

\emergencystretch=\maxdimen
\begin{document}

\baselineskip 20 true pt plus3pt minus3pt
\renewcommand\arraystretch{1.5}
\renewcommand{\baselinestretch}{1.15}
\newcommand{\supercite}[1]{\textsuperscript{\cite{#1}}}
\newcommand{\upcite}[1]{\textsuperscript{\textsuperscript{\cite{#1}}}}
\newcommand{\reff}[1]{(\ref{#1})}

\vspace*{0.05 in}
\begin{center}
\bf \Large Shear-driven dynamics of surfactant-laden droplets on rough substrates
\end{center}
\begin{center}
\large Ninad V. Mhatre and Satish Kumar\footnote[2]{Email address for correspondence: kumar030@umn.edu}\\
\normalsize
Department of Chemical Engineering and Materials Science, University of Minnesota, Minneapolis, MN 55455, USA\\
\normalsize
\end{center}

\begin{abstract}
\noindent

The depinning of liquid droplets due to flow of a surrounding immiscible fluid plays a crucial role in applications such as enhanced oil recovery, surface cleaning, and crossflow emulsification. Although surfactants are often present in these systems, the role of Marangoni stresses on droplet depinning by an external flow remains unclear. To address this, we develop a lubrication-theory-based model for a thin Newtonian droplet laden with insoluble surfactant on a substrate with Gaussian-shaped defects which are used to account for the effects of surface roughness. The droplet is surrounded by a surfactant-free immiscible Newtonian fluid in a long, narrow rectangular channel, with flow driven by an applied pressure gradient. Using a precursor-film/disjoining-pressure approach for contact-line motion, we derive nonlinear evolution equations for the droplet thickness and interfacial surfactant concentration, which are solved numerically. The pressure gradient transports surfactant from the receding to the advancing contact line, generating a Marangoni flow opposing the pressure-driven flow. This reduces the net shear force on the droplet, leading to depinning at a higher critical pressure gradient. These findings reveal a previously unexamined regime in which interfacial Marangoni stresses, rather than uniform interfacial-tension reduction, govern the critical flow rate. The results provide a mechanistic basis for using surfactant-concentration gradients as a tunable handle to control droplet motion on rough substrates.

\vspace{0.5cm}

\noindent
\end{abstract}
\pagebreak
\section{Introduction} \label{sec:Intro}
The depinning of liquid droplets due to flow of a surrounding immiscible fluid plays a crucial role in practical applications. Enhanced oil recovery uses a water flow to depin oil droplets inside rock crevices and depleted reservoirs \cite{howe2015visualising, delshad2009modeling}. Several surface-cleaning methods apply a fluid flow to depin and remove contaminants in the form of liquid droplets. \cite{onaizi2009rapid, chelazzi2020use, landel2021fluid}  Crossflow emulsification involves a flow of a continuous phase to depin droplets of a dispersed phase from membrane pores \cite{charcosset2004membrane,salama2022estimation}. In these applications, surfactants may be present as impurities or purposefully added, and their presence can significantly affect depinning.

One of the simplest ways to rationalize droplet depinning is with a force-balance model, in which the drag force exerted on the droplet by the surrounding fluid drives depinning and the surface-tension force along the droplet contact line resists depinning. Above a critical flow rate, the drag force exceeds the surface-tension force and the droplet depins \cite{dussan1987drops, mahe1987pure, mahe1988rough,basu1997model,elsherbini2006retention,gupta2008deformation,antonini2009general,milne2009drop,seevaratnam2010laminar,fan2011displacement,madani2014oil,roisman2015dislodging,lu2019deformation}. The drag force, $F_s$, is estimated by assuming a static droplet with a spherical-cap shape and Stokes-like drag law.  It has the form $F_s \sim \mu_s R_0 v_s$, where $\mu_s$ is the viscosity of the surrounding fluid, $R_0$ is the radius of the droplet contact line, and $v_s$ is a characteristic velocity of the surrounding fluid. The surface-tension force, $F_{surf}$, is calculated by assuming a stationary and circular contact line such that the contact angle in its entire advancing half is equal to the advancing contact angle, $\theta_{acl}$, and the contact angle in its entire receding half is equal to the receding contact angle, $\theta_{rcl}$. It has the form $F_{surf} \sim \sigma R_0 (\cos{\theta_{rcl}} - \cos{\theta_{acl}})$, where $\sigma$ is the interfacial tension between the droplet and the surrounding fluid.

Surfactants can affect this force balance in multiple and competing ways.  First, surfactants will generally lower the interfacial tension, which would reduce the surface-tension force.  Second, by lowering the interfacial tension, the equilibrium contact angle of the droplet will decrease (via Young's equation), causing the droplet to spread more.  This increase in the contact-line radius would increase the surface-tension force.  Third, the surfactants could adsorb to the solid substrate and affect the contact-angle hysteresis, which is reflected in the difference between $\theta_{acl}$ and $\theta_{rcl}$.  Fourth, changes to the droplet shape, which can be caused by any of the above three factors, will affect the drag force on the droplet.  Finally, gradients in surfactant concentration along the droplet interface will generate Marangoni stresses, which can also affect droplet shape.  Notably, Marangoni stresses do not explicitly appear in the force-balance model, and the objective of the present work is to develop a model to understand how they affect depinning in the limiting case of thin droplets and insoluble surfactants. 

Experiments aimed at understanding how surfactants affect the critical flow rate at which droplets depin typically consider an oil droplet in the presence of a flowing surfactant-laden aqueous solution \cite{rowe2002oil,thoreau_physico-chemical_2006,freville2014effect}, although surfactants can be introduced into the oil droplet as well \cite{mahe1988contaminated}. Depending on the experimental conditions, it is found that surfactants can either increase or decrease the critical flow rate (or shear rate).  The results are typically rationalized using the force-balance model discussed above, and are consistent with the observation that  surfactants can affect the force balance in multiple and competing ways.  One study hypothesized that Marangoni stresses can promote depinning \cite{thoreau_physico-chemical_2006}, although a detailed description of the mechanism by which this occurs was not provided.

At a microscopic level, contact-line pinning occurs due to surface heterogeneities, which can be either topographical or chemical in nature.  At a macroscopic level, these heterogeneities are reflected implicitly in contact-angle hysteresis, the difference between advancing and receding contact angles.  While force-balance models are among the simplest to describe the depinning of contact lines by an external flow, they do not provide insight into the mechanics of depinning near explicit surface heterogeneities, which is not only of fundamental interest but also potentially useful information for designing heterogeneous surfaces to control droplet pinning and depinning.  Moreover, force-balance models cannot predict steady or transient droplet shapes, and require knowledge of $\theta_{acl}$ and $\theta_{rcl}$ as well as the radius of the droplet contact line ($R_0$) at the point of depinning to predict a critical flow rate for depinning.  Finally, these models also make assumptions about the droplet having a spherical-cap shape, circular contact-line, and constant values of $\theta_{acl}$ and $\theta_{rcl}$ along the contact line, none of which may be true in general.

To begin to address these limitations, in our prior work we developed a lubrication-theory-based model of droplet depinning on rough substrates \cite{mhatre2024shear}. The model considers thin two-dimensional droplets (liquid ridges) in a rectangular channel surrounded by another immiscible fluid with flow driven by a constant pressure gradient.  Surface roughness is incorporated by considering two Gaussian-shaped bumps on each side of the droplet, and contact-line motion is described with a precursor-film/disjoining-pressure approach.  Numerical solutions of the resulting nonlinear evolution equation describing the droplet thickness as a function of space and time reveal that the droplet remains pinned below a critical pressure gradient.  Above this value, the droplet depins and slides along the surface at a constant velocity.  Under some conditions, small residual droplets can be left behind at the defect after depinning.

The pinning/depinning transition can be understood by considering a balance between viscous and surface-tension forces, or via a balance between capillary-pressure
gradients and disjoining-pressure gradients \cite{mhatre2024shear}. It is found that the receding and
advancing contact lines always pin at the points on the defects that have the maximum
negative slope because this maximizes the surface-tension force acting on the droplet. Variations in the critical pressure gradient arise from the way different parameters modify these forces.  The critical pressure gradient decreases with increasing droplet volume, surface wettability, and defect width, and increases with increasing defect height. As the viscosity of the surrounding fluid increases, the critical pressure gradient first decreases, reaches a minimum, and then increases again.  Many of the observations can be rationalized using simple analytical models and are qualitatively consistent with experimental observations.  The model can readily be extended to consider three-dimensional effects, additional topographical defects, chemical heterogeneities, and more complicated defect shapes.  The model complements prior related work concerning droplet pinning/depinning near chemical heterogeneities and on substrates with topography variations \cite{joanny1984model,raphael1989dynamics,joanny1990motion,beltrame2009depinning,beltrame2011rayleigh,thiele2006driven,savva2013droplet,park2017droplet,shang2019droplets,li2023sliding}, as well as numerical simulations of droplet depinning by flow of a surrounding fluid that do not explicitly incorporate surface heterogeneities \cite{schleizer_displacement_1999,ding2008onset,ding2010sliding,deng2022modeling}.

The goal of the present work is to extend the model of Mhatre and Kumar\cite{mhatre2024shear} to understand the influence of Marangoni stresses on droplet depinning by a shear flow of an adjacent immiscible fluid.  In addition to being of fundamental interest, a better understanding of the role of Marangoni stresses on droplet depinning by an external flow may be helpful for choosing surfactants for the applications mentioned above.  Moreover, the model we develop allows us to begin addressing the hypothesis from prior work\cite{thoreau_physico-chemical_2006} that claims Marangoni stresses can promote depinning.

To isolate basic physical mechanisms, we focus on the case where the droplets are thin and the surfactants are insoluble, i.e., localized to the liquid-liquid interface. The model formulation is discussed in \S\ref{sec:Model_formulation}, followed by an investigation of droplet dynamics on smooth substrates in \S\ref{sec:smooth substrate}.  This lays a foundation for a study of droplet dynamics on a substrate with topographical defects in \S\ref{sec:rough substrate} and \S\ref{sec:surfactant influence}.  Concluding remarks are given in \S\ref{sec:conclusions}.

\section{Model formulation} \label{sec:Model_formulation}
\FloatBarrier
\begin{figure}[t]
\centering
\begin{subfigure}{0.58\textwidth}
\includegraphics[width=\linewidth]{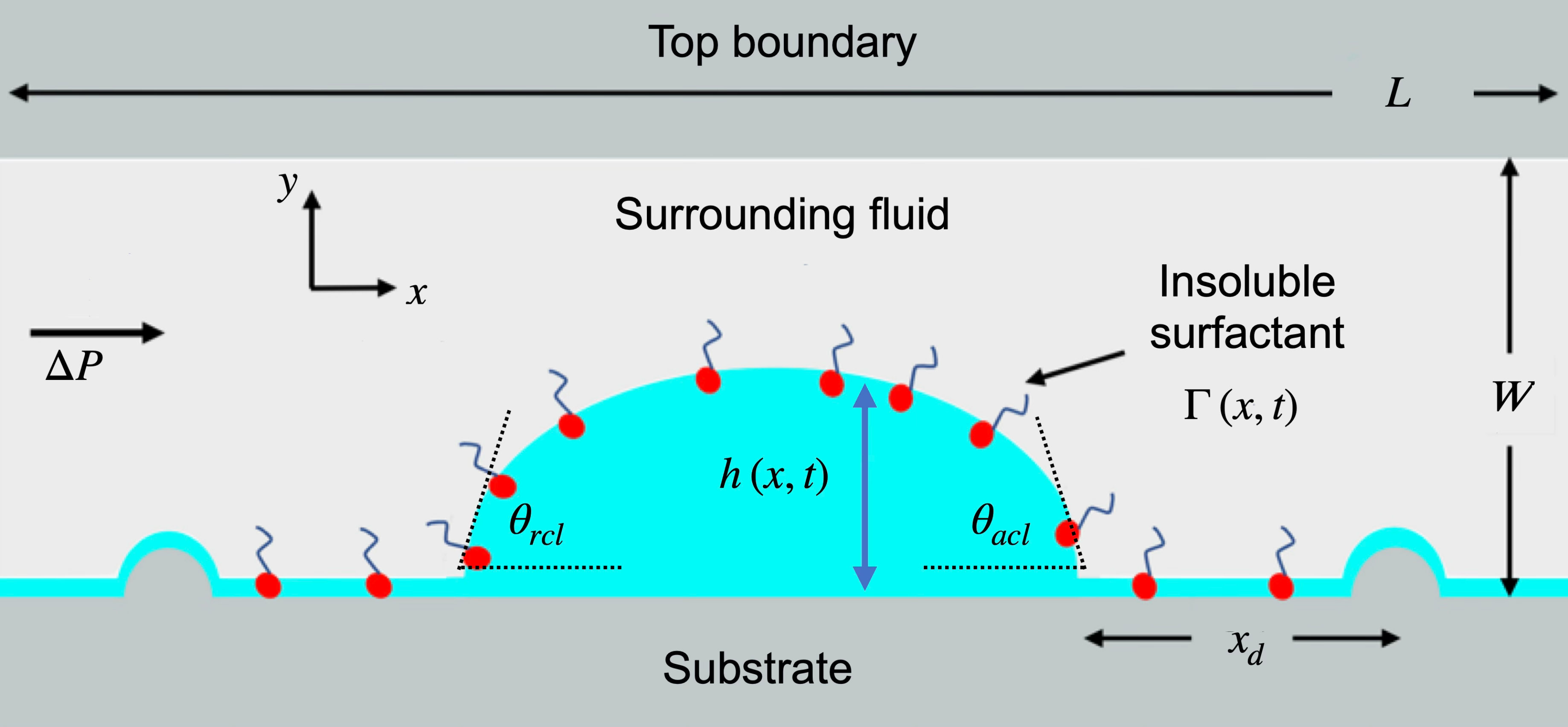}
\caption{}
\label{fig:schematic1}
\end{subfigure}
\hfill
\begin{subfigure}{0.38\textwidth}
\includegraphics[width=\linewidth]{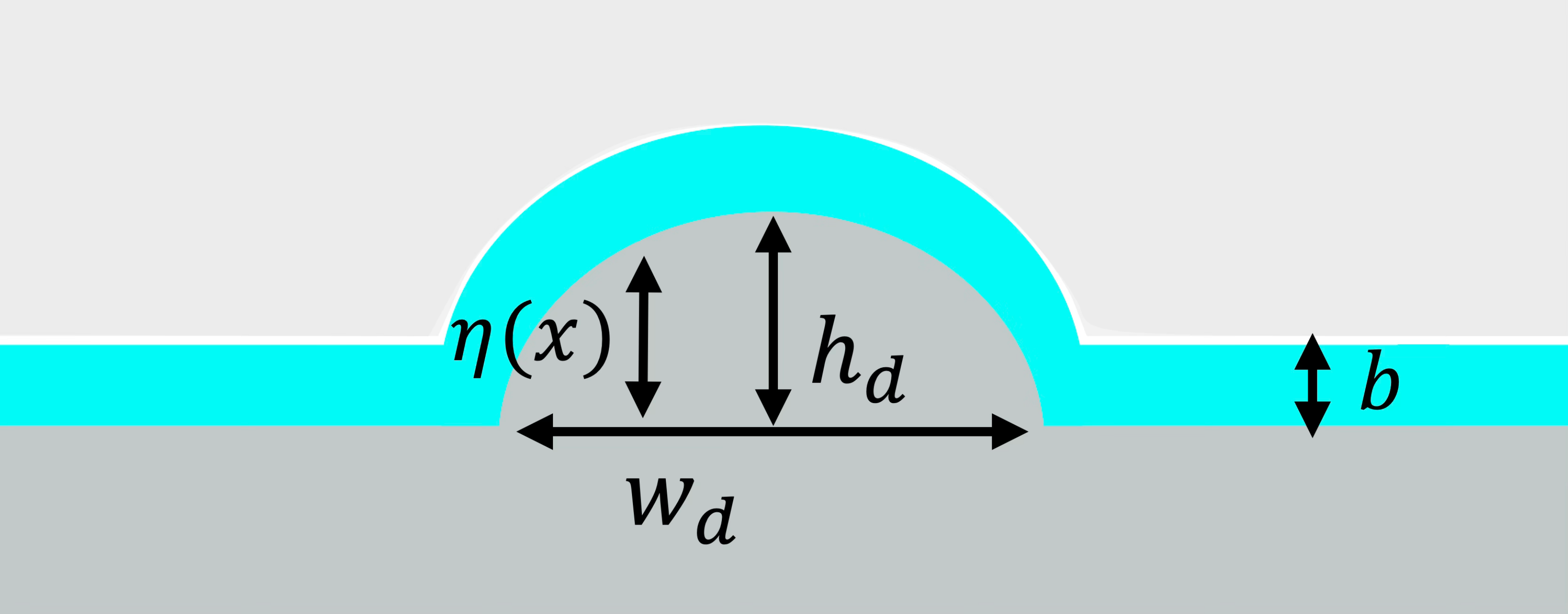}
\caption{}
\label{fig:schematic2}
\end{subfigure}
\caption{(\subref{fig:schematic1}) Problem schematic. (\subref{fig:schematic2}) Enlarged view of a substrate defect.}
\label{fig:fig1}
\end{figure}

\subsection{Problem geometry} \label{subsec: problem geometry}

Figure \ref{fig:schematic1} shows a problem schematic, where a rectangular channel contains a two-dimensional Newtonian droplet surrounded by another immiscible Newtonian fluid with surfactant present only at the droplet interface.  Here, $L$ is the channel length, $W$ is the channel width, $h(x,t)$ is the droplet height, $\Gamma(x,t)$ is the surfactant concentration, $\theta_{rcl}$ is the apparent receding contact angle, and $\theta_{acl}$ is the apparent advancing contact angle. The horizontal direction is represented by $x$, the vertical direction by $y$, and time by $t$. A thin droplet is considered here such that its height is significantly smaller than its maximum width, and the channel is assumed to be long and narrow such that $W \ll L$. This permits the use of the lubrication approximation to simplify the governing equations. A negative pressure gradient (technically, pressure change) of magnitude $\Delta P$ is applied in the channel to drive flow from left to right. Although a two-dimensional droplet is actually a ridge, for simplicity we will refer to it as a droplet throughout this paper.

We incorporate surface roughness on the substrate (channel bottom) in the form of two bump-type defects present at a distance $x_d$ in front of and behind the droplet (figure \ref{fig:schematic1}). An enlarged view of one defect is shown in figure \ref{fig:schematic2}, where $h_d$ is the maximum defect height and $w_d$ is the maximum defect width. Following prior work on related problems \cite{espin2015droplet,pham2017drying,pham2019imbibition,mhatre2024shear,mhatre_pinningdepinning_2024}, we describe the shape of the bump using a Gaussian function as it is simple to implement computationally and yields droplet dynamics that are qualitatively consistent with experimental observations.  The bump shape is described as $\eta(x) = h_d \{$exp$(-(x - x_{c1})^2/(2 w_d^2)) + $exp$(-(x - x_{c2})^2/(2 w_d^2))\}$, where $x_{c1}$ is the center of the defect on the left and $x_{c2}$ is the center of the defect on the right.  Although in practice substrates can have multiple defects with a distribution of sizes, shapes, and locations, the limiting case we consider here is sufficient for obtaining basic physical understanding.

We note that for the lubrication approximation to formally apply, we need $h_{max} \ll D$, with  $h_{max} \sim W$ and $D \sim L$ at most, where $D$ is the droplet diameter\cite{mhatre2024shear}.  A similar restriction applies to the bump, and both the droplet interface and bump need to have small slopes.  However, even outside these limits, lubrication theory can work surprisingly well, as evidenced by comparison to numerical simulations of the full governing equations in related problems (e.g., Ref. \citeonline{zhou2012two}).  We note that the case of thin droplets corresponds to small equilibrium contact angles and is relevant to cases where the droplet strongly wets the substrate.   

\subsection{Scaling and evolution equations} \label{subsec: governing equations}

We choose characteristic scales similar to those in our prior work \cite{mhatre2024shear}. The horizontal and vertical distances are non-dimensionalized with $L$ and $W$, respectively. All stresses are non-dimensionalized with a characteristic capillary pressure $W\sigma_m/L^2$, where $\sigma_m$ is the interfacial tension corresponding to the mean surfactant concentration $\Gamma_m$ at the interface. The horizontal velocity is non-dimensionalized with a capillary spreading speed $u^* = \epsilon^3\sigma_m/3\mu^d$, where $\mu^d$ is the droplet viscosity, the vertical velocity is non-dimensionalized with $\epsilon u^*$, and time is non-dimensionalized with $L/u^*$. The surfactant concentration is non-dimensionalized with $\Gamma_m$ and the dimensionless surface tension is $\sigma = (\sigma' - \sigma_m)/(\sigma_0 - \sigma_m)$, where $\sigma_0$ is the surface tension corresponding to zero surfactant concentration and $\sigma'$ is used to denote the dimensional surface tension.  Unless otherwise noted, all variables from now on are in dimensionless form.

Motivated by prior work on droplet motion on solid substrates, we use a precursor-film/disjoining-pressure approach to model contact-line dynamics. \cite{schwartz1998simulation, schwartz1998hysteretic, espin2015droplet, espin2017droplet, mhatre_pinningdepinning_2024, mhatre2024shear} Unlike approaches that apply a slip condition on the substrate, here the contact-line position is not an  
additional variable but is extracted from the droplet height profile, as are the advancing and receding contact angles (see \S\ref{subsec:contact angles}).  This approach makes the equations less complicated to solve\cite{savva2011dynamics} and assumes the presence of a thin precursor film of thickness $b$ on the entire substrate (Figure \ref{fig:fig1}).  As in prior work, we use a two-term disjoining pressure, $\Pi$:
\begin{gather} 
\Pi = A \left[\left(\frac{b}{h}\right)^n - \left(\frac{b}{h}\right)^m\right], \label{eq: disj press} 
\end{gather}
where $A$ is the dimensionless Hamaker constant and $h$ is the droplet thickness. The term with exponent $n$ corresponds to repulsive intermolecular forces, while the term with exponent $m$ represents attractive forces. We use $n=3$ and $m=2$, as these values have been shown to provide a qualitatively accurate description of contact-line dynamics with reasonable computational efficiency. \cite{espin2015droplet, espin2017droplet, mhatre_pinningdepinning_2024, mhatre2024shear} Furthermore, this model can be extended to account for chemical heterogeneity on the substrate by allowing $A$ to vary spatially. \cite{schwartz1998simulation, schwartz1998hysteretic, kalpathy2012thin}

An equilibrium contact angle of magnitude $\theta_{eq}$ is determined by the two-term disjoining pressure, which the droplet reaches in the absence of an applied pressure gradient:\cite{schwartz1998simulation}
\begin{gather}
\theta_{eq} = \sqrt{\frac{2(n-m)A b}{(n-1)(m-1)}}. \label{eq: equilibrium contact angle}
\end{gather}
Here, $\theta_{eq}$ is the scaled equilibrium contact angle which is related to the actual lab-frame equilibrium contact angle by $\theta_{eq, lab} = \epsilon \theta_{eq}$. All the angles discussed hereafter are scaled angles. We assume that the surfactant concentration is dilute enough to impose a linear equation of state:\cite{jensen1992insoluble, matar_rupture_2004, craster2009dynamics}
\begin{gather}
\sigma = 1 - \Gamma. \label{eq:linear EOS}
\end{gather}

The equations governing conservation of total mass, linear momentum, and surfactant along with their boundary conditions can be reduced into two coupled nonlinear evolution equations for the droplet height and surfactant concentration.  As the derivation is similar to that in our prior work\cite{mhatre2024shear} and in previous works involving thin films of insoluble surfactant\cite{oron1997long,craster2009dynamics}, we simply present the final results,
\begin{multline}
\frac{\partial h}{\partial t} = \frac{\partial}{\partial x} \biggl[ \frac{(H-1)^3 h^3 (1+H(\mu_r-1) -\mu_r\eta)}{g(H,\mu_r, \eta)} \left(\frac{\partial \Pi}{\partial x} + \epsilon^2 M \frac{\partial \Gamma}{\partial x}\frac{\partial^2H}{\partial x^2} + (\epsilon^2 M (\Gamma-1) - 1)\frac{\partial^3 H}{\partial x^3}\right) \\  - \frac{3M h^2 (H-1)^2 (\mu_r \eta^2 - 2H\mu_r\eta + H(2+H(\mu_r-1))-1)}{2g(H,\mu_r,\eta)} \frac{\partial \Gamma}{\partial x} \\ + \Delta P \frac{h^2 (3+H^2(\mu_r-1)-2H(1+(\mu_r-2)\eta + \eta(\mu_r \eta - 4))}{4 g(H,\mu_r, \eta)} \biggr], \label{eq:height evolution equation}
\end{multline}

\begin{multline}
\frac{\partial \Gamma}{\partial t} = \frac{1}{Pe} \frac{\partial^2 \Gamma}{\partial x^2} +  \frac{\partial}{\partial x} \biggl[ \Gamma \biggl(\Delta P \frac{3 h (H-1)(\eta-1)}{2 g(H,\mu_r, \eta)} \\ + \frac{3(H-1)^2 h^2 (\mu_r \eta^2-2H\mu_r\eta+H(2+H(\mu_r-1))-1)}{2g(H,\mu_r, \eta)} \biggl( \frac{\partial \Pi}{\partial x} + \epsilon^2 M \frac{\partial \Gamma}{\partial x}\frac{\partial^2H}{\partial x^2} \\ + (\epsilon^2 M (\Gamma-1) - 1)\frac{\partial^3 H}{\partial x^3}\biggr) \\ - M \frac{\partial \Gamma}{\partial x} \frac{3h(H-1)(1+H^3(\mu_r-1) - \mu_r \eta^3 + 3H^2(1-\mu_r\eta) + 3H(\mu_r \eta^2-1))}{g(H,\mu_r,\eta)} \biggr) \biggr], \label{eq:surfactant evolution equation}
\end{multline}
where $H(x,t) = h(x,t)+\eta(x)$ and
\begin{multline}
g(H,\mu_r,\eta) = 1+H^4(\mu_r-1)^2 - 4H^3(\mu_r-1)(\mu_r\eta-1) \\ +6H^2(\mu_r-1)(\mu_r\eta^2-1) - 4H(\mu_r-1)(\mu_r\eta^3-1) + \mu_r\eta(-4 + \eta(6+\eta(\mu_r\eta - 4))). \label{eq: evo_eqn2}
\end{multline}
Here, $\mu_r = \mu^s / \mu^d$ is the viscosity ratio, with $\mu^s$ and $\mu^d$ representing the viscosity of the surrounding fluid and droplet, respectively.  Our derivation assumes that the droplet is thin enough so that gravitational forces are negligible relative to surface-tension forces, and as a consequence, a density ratio does not appear in the problem.

Two important dimensionless parameters appear in
(\ref{eq:height evolution equation}) and (\ref{eq:surfactant evolution equation}). 
 The Peclet number $Pe = (u^*L)/D_s = (\epsilon^3 \sigma_m L)/(3 \mu_d D_s)$, where $D_s$ is the diffusion coefficient of the surfactant along the interface, and represents the ratio of the rate of convection to the rate of diffusion of surfactant along the interface.  The Marangoni number $M = [(-\partial \sigma'/\partial \Gamma') \Gamma_m]/\epsilon^2 \sigma_m$, where $-\partial \sigma'/\partial \Gamma'$ is the rate of change of surface tension with the surfactant concentration, with $\Gamma'$ being the dimensional surfactant concentration.  Here, $M$ provides a measure of forces arising from surface-tension gradients to those arising from the mean surface tension.  

 It is instructive to point out that in our problem formulation, the Marangoni number appears in both the tangential and normal stress balances, which in the lubrication limit are, respectively, 
 \begin{gather}
 \partial u^d/\partial y = \mu_r (\partial u^s/\partial y) - 3M(\partial \Gamma/\partial x), \label{eq: tangential stress balance} \\    
 p^d - p^s = -(1-\epsilon^2 \sigma M)H_{xx}-\Pi, \label{eq: normal stress balance}
 \end{gather}
where $u$ is the horizontal velocity and $p$ is the pressure, with the superscripts `$d$' and `$s$' indicating droplet and surrounding fluid, respectively.  Equation (\ref{eq: tangential stress balance}) shows that for thin droplets, vertical gradients in the horizontal velocity are coupled via the Marangoni number to horizontal gradients in interfacial concentration of surfactant.  Equation (\ref{eq: normal stress balance}) shows that the normal stress balance can be affected by $M$ via a reduction in the interfacial tension.  However, this effect only becomes important if $\epsilon^2 M \sim 1$.   We retain this term to explore cases where this term has a significant effect on the results.  

\subsection{Numerical solution} \label{subsec: governing equations}

The length of the computational domain is set to $L^*$, and the interface between the precursor film and the surrounding fluid is required to be flat at the left and right ends by applying the following conditions:
\begin{gather}
h(x = 0, t) = b, \; \; \; \;\; \hfill h_x (x = 0, t) = 0, \label{eq: BC at x = 0} \\
h(x=L^*, t) = b, \; \; \; \; \; \hfill h_x (x = L^*, t) = 0. \label{eq: BC at x = L}
\end{gather}
No-flux conditions on the surfactant concentration are applied at the ends of the domain:
\begin{gather}
\Gamma_x(x = 0, t) = 0 \label{eq: BC at x = 0 for gamma} \\
\Gamma_x(x=L^*, t) = 0. \label{eq: BC at x = L for gamma}
\end{gather}

We define the initial droplet shape for a given dimensionless droplet volume, $v_0$, by specifying a fourth-order polynomial which satisfies (\ref{eq: BC at x = 0}) and (\ref{eq: BC at x = L}). The droplet height and surfactant concentration evolution equations, (\ref{eq:height evolution equation}) and (\ref{eq:surfactant evolution equation}), are discretized using a fully implicit-centered fourth-order finite-difference scheme with $800$-$1000$ nodes per unit length of the computational domain, and the MATLAB built-in solver ode15s is used for time integration.

We use $L^*=6$ or $L^*=9$ to obtain results that are independent of the length of the computational domain. We set $b=0.001$ as it recovers Tanner's spreading law for a two-dimensional droplet at a reasonable computational cost \cite{mhatre2024shear}. We have already conducted a systematic parametric study to understand the influence of $\mu_r$, $v_0$, $\theta_{eq}$, $h_d$, and $w_d$ on droplet dynamics in our prior work \cite{mhatre2024shear}, as discussed in \S\ref{sec:Intro}, and we fix these quantities for the calculations presented in this paper to isolate the effects of $M$ and $Pe$. We use $\mu_r = 0.01$, $v_0 = 0.2$, $\theta_{eq}=10^{\circ}$ ($A=10^5$ from (\ref{eq: equilibrium contact angle})), $h_d = 0.05 h_{max}$, and $w_d = 2.5h_d$, where $h_{max}$ is the maximum droplet height when it reaches $\theta_{eq}$ in the absence of an applied pressure gradient.  We fix $\epsilon = 0.01$, and based on the values listed in Table \ref{tab:dimensional_ranges_MP}, we vary $M$ from 1 to 100 and $Pe$ from 10 to 1000.

\begin{table}[t]
\centering
\caption{Representative ranges of dimensional quantities used to obtain $M$ and $Pe$ values consistent with those used in the present calculations.}
\label{tab:dimensional_ranges_MP}
\renewcommand{\arraystretch}{1.3}
\begin{tabular}{|c|c|}
\hline
Variable & Range of values \\
\hline
Channel length, $L$ (m) & $\sim 10^{-2}$--$10^{-1}$ \\
\hline
Channel width, $W$ (m) & $\sim 10^{-4}$--$10^{-3}$ \\
\hline
Interfacial tension at mean surfactant concentration, $\sigma_m$ (N/m) & $\sim 10^{-2}$ \\
\hline
Droplet viscosity, $\mu^d$ (Pa$\cdot$s) & $\sim 10^{-3}$--$10^{-1}$ \\
\hline
Interfacial surfactant diffusivity, $D_s$ (m$^2$/s) & $\sim 10^{-10}$--$10^{-8}$ \\
\hline
Characteristic surface-tension change, $\left(-\partial \sigma'/\partial \Gamma'\right)\Gamma_m$ (N/m) & $\sim 10^{-6}$--$10^{-3}$ \\
\hline
\end{tabular}
\end{table}


\subsection{Contact angles and contact lines} \label{subsec:contact angles}
\begin{figure}[t]
\centering
\begin{subfigure}{0.49643\textwidth}
\includegraphics[width=\linewidth]{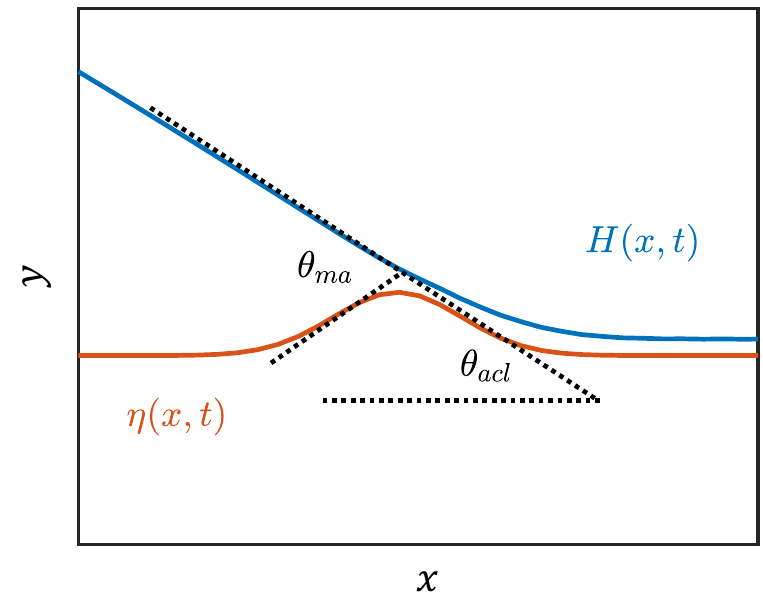}
\caption{}
\label{fig:meso_angle_acl}
\end{subfigure}
\hfill
\begin{subfigure}{0.49643\textwidth}
\includegraphics[width=\linewidth]{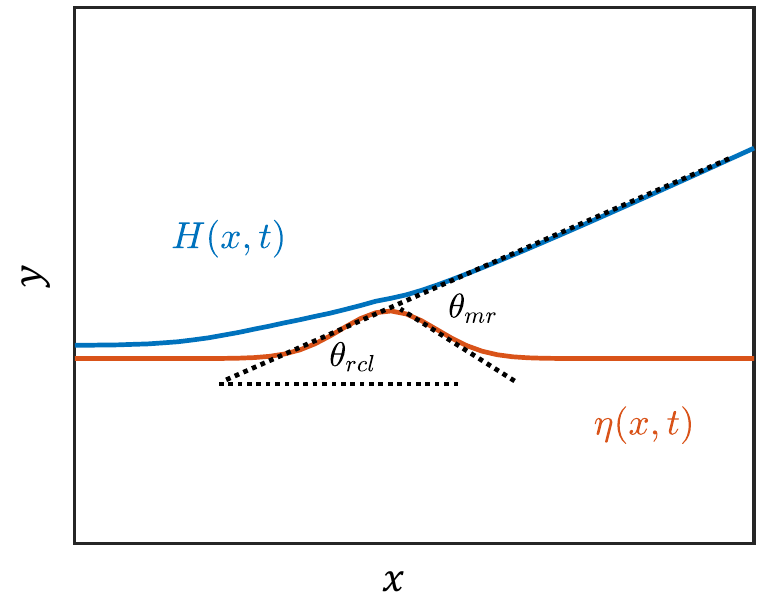}
\caption{}
\label{fig:meso_angle_rcl}
\end{subfigure}
\caption{(\subref{fig:meso_angle_acl}) Contact-line region when the advancing contact line slides over a defect, where $\eta$ shows the substrate shape, $H$ shows the interface shape, $\theta_{acl}$ is the apparent advancing contact angle, and $\theta_{ma}$ is the largest mesoscopic angle on the advancing half of the droplet. (\subref{fig:meso_angle_rcl}) Contact-line region when the receding contact line slides over a defect, where $\theta_{rcl}$ is the apparent receding contact angle, and $\theta_{mr}$ is the largest mesoscopic angle on the receding half of the droplet.}
\label{fig:fig2}
\end{figure}

For a smooth substrate, the apparent advancing ($\theta_{acl}$) and receding ($\theta_{rcl}$) contact angles (Figure \ref{fig:schematic1}) are defined as the largest angles between the substrate and the tangents to the droplet interface, on the advancing and receding sides of the droplet, respectively. The advancing and receding contact-line locations, $x_{acl}$ and $x_{rcl}$, are defined as the points of intersection between the tangents corresponding to the apparent contact angles and the substrate. When there is no applied pressure gradient in the channel, $\theta_{acl} = \theta_{rcl}$. If $\theta_{acl} > \theta_{eq}$, the droplet spreads until $\theta_{acl} = \theta_{eq}$, and if $\theta_{acl} < \theta_{eq}$ the droplet retracts until $\theta_{acl} = \theta_{eq}$.

A mesoscopic contact angle $\theta_m$ is defined for substrates with topographical defects \cite{espin2015droplet, park2017droplet},
\begin{gather}
{\text{tan}}(\theta_m) = \frac{h_x}{1+(h_x + \eta_x)h_x}. \label{eq: mesoscopic angle definition}
\end{gather}
Here, $\theta_{acl}$ and $\theta_{rcl}$ are obtained by finding the points on the droplet interface where $\theta_m$ is the largest on the advancing ($\theta_{ma}$ in figure \ref{fig:meso_angle_acl}) and receding  ($\theta_{mr}$ in figure \ref{fig:meso_angle_rcl}) sides of the droplet, respectively, and then extrapolating the tangents at these points to the substrate. 




\section{Droplet dynamics on a smooth substrate} \label{sec:smooth substrate}

\begin{figure}[t]
\centering
\begin{subfigure}{0.35\textwidth}
\includegraphics[width=\linewidth]{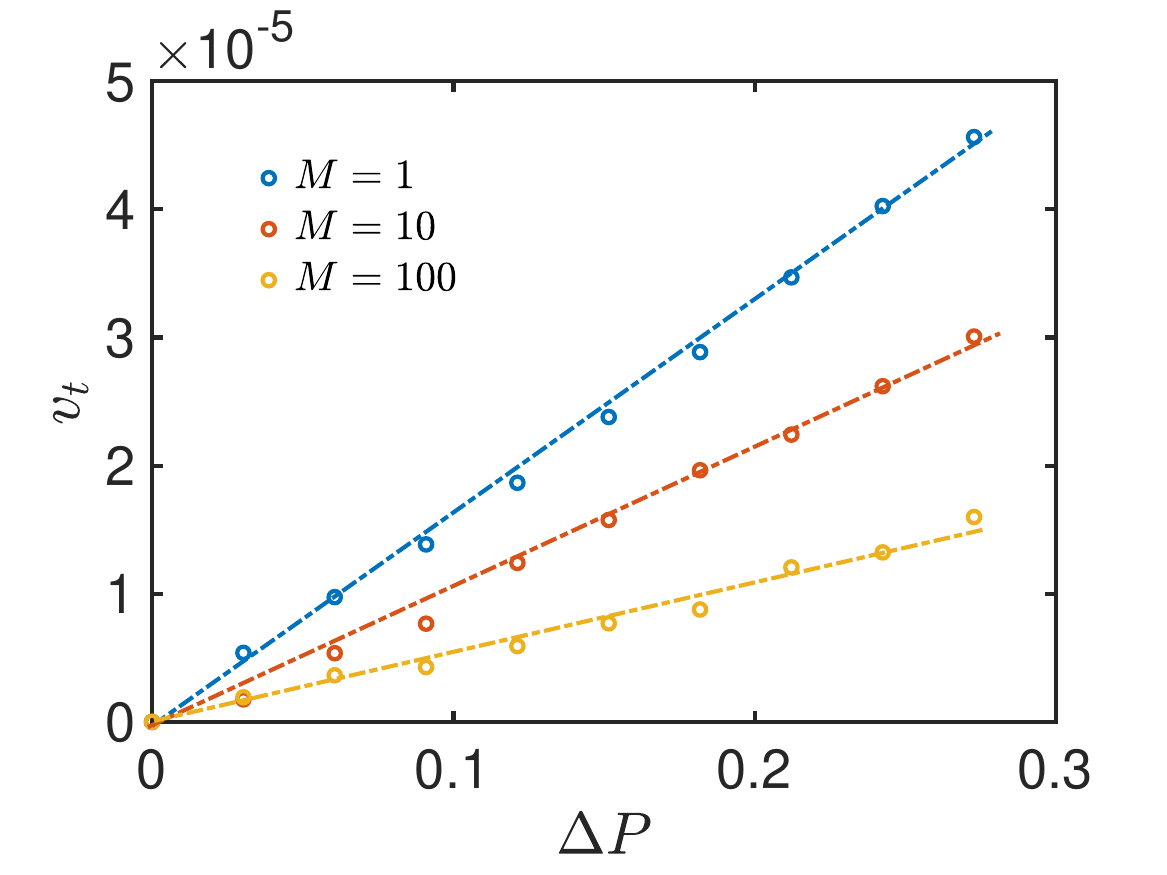}
\caption{}
\label{fig:vt vs dP variation with M}
\end{subfigure}
\begin{subfigure}{0.35\textwidth}
\includegraphics[width=\linewidth]{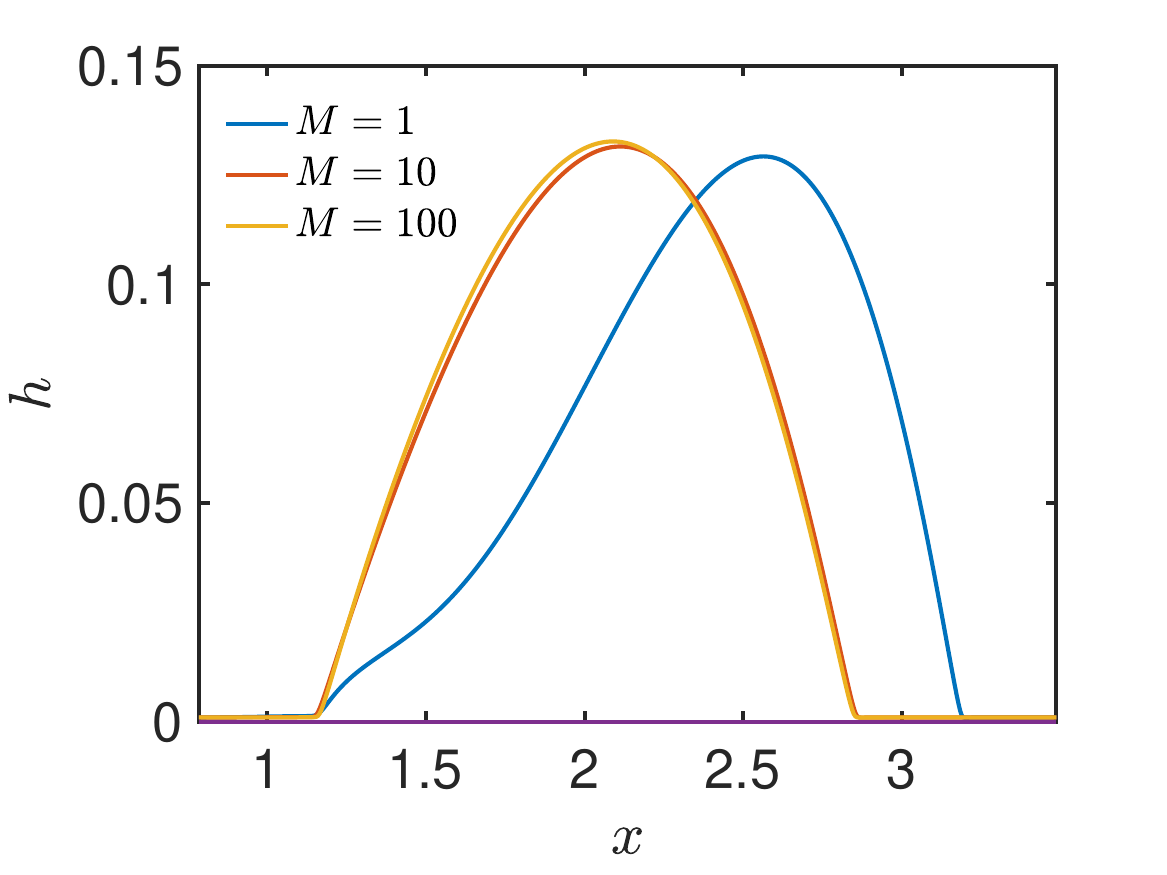}
\caption{}
\label{fig:droplet profiles on smooth for diff M}
\end{subfigure}
\begin{subfigure}{0.35\textwidth}
\includegraphics[width=\linewidth]{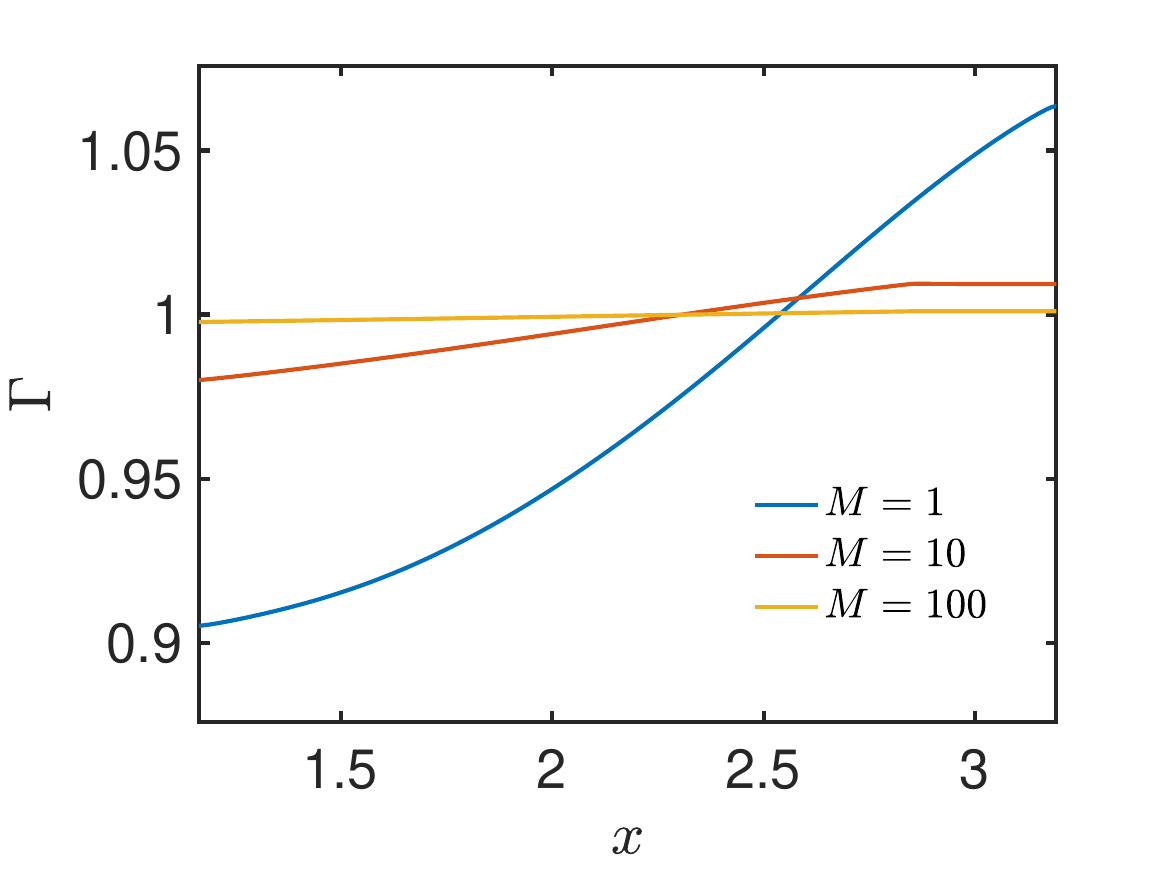}
\caption{}
\label{fig:surf conc on smooth for diff M}
\end{subfigure}
\begin{subfigure}{0.35\textwidth}
\includegraphics[width=\linewidth]{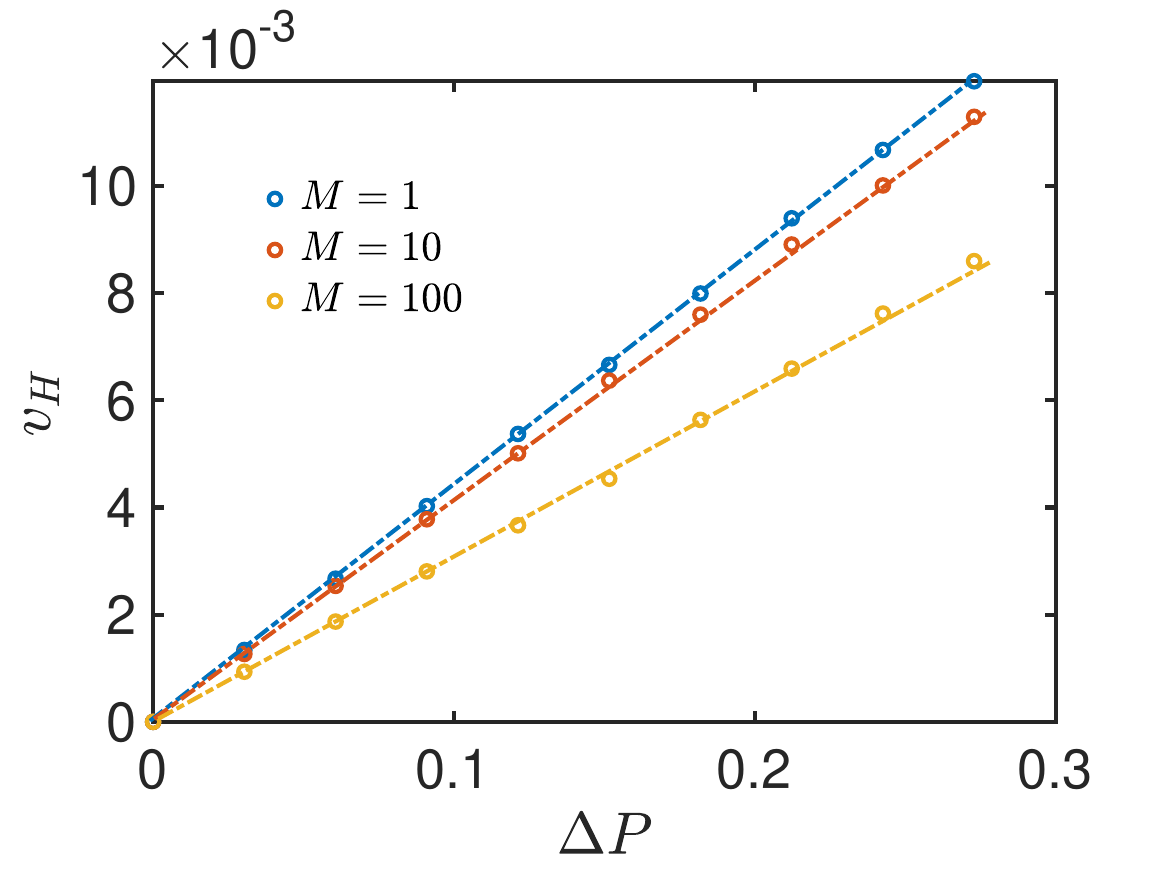}
\caption{}
\label{fig:vt analytical vs dP variation with M}
\end{subfigure}
\caption{(\subref{fig:vt vs dP variation with M}) $v_t$ vs. $\Delta P$ for different $M$. The open circles show numerical calculations and the dashed lines show linear fits. (\subref{fig:droplet profiles on smooth for diff M}) Steady droplet shapes for different $M$ with $\Delta P = 0.03$. The droplets have been shifted such that their receding contact lines coincide. (\subref{fig:surf conc on smooth for diff M}) Steady surfactant-concentration profiles ($\Gamma$ vs. $x$) for different $M$ with $\Delta P = 0.03$. (\subref{fig:vt analytical vs dP variation with M}) $v_H$ vs. $\Delta P$ for different $M$. The open circles show numerical calculations and the dashed lines show linear fits. The other parameters are $Pe = 1000$, $L^*=6$, $\mu_r = 0.01$, $v_0 = 0.2$, $\theta_{eq}=10^{\circ}$ ($A=10^5$), and $b=0.001$}
\label{fig:fig3}
\end{figure}

\begin{figure}[t]
\centering
\begin{subfigure}{0.348\textwidth}
\includegraphics[width=\linewidth]{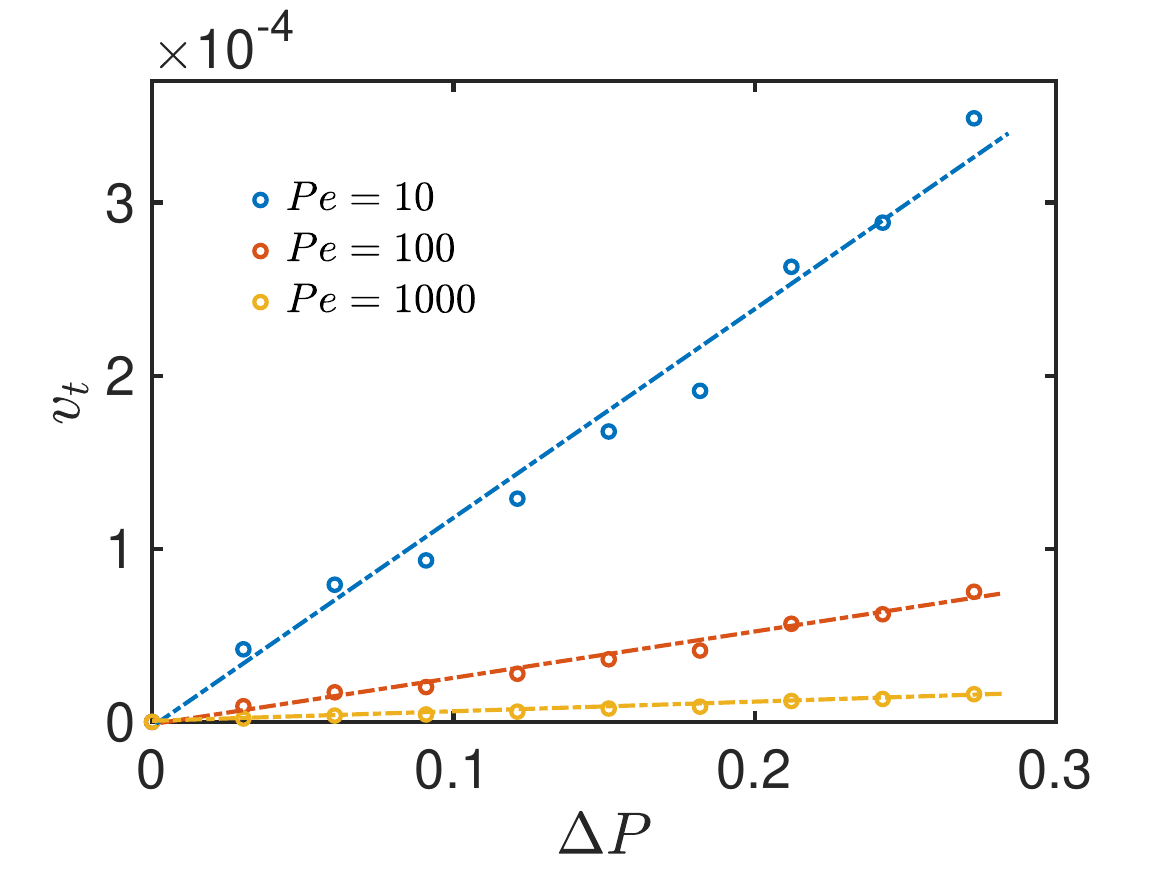}
\caption{}
\label{fig:vt vs dP variation with Pe}
\end{subfigure}
\begin{subfigure}{0.348\textwidth}
\includegraphics[width=\linewidth]{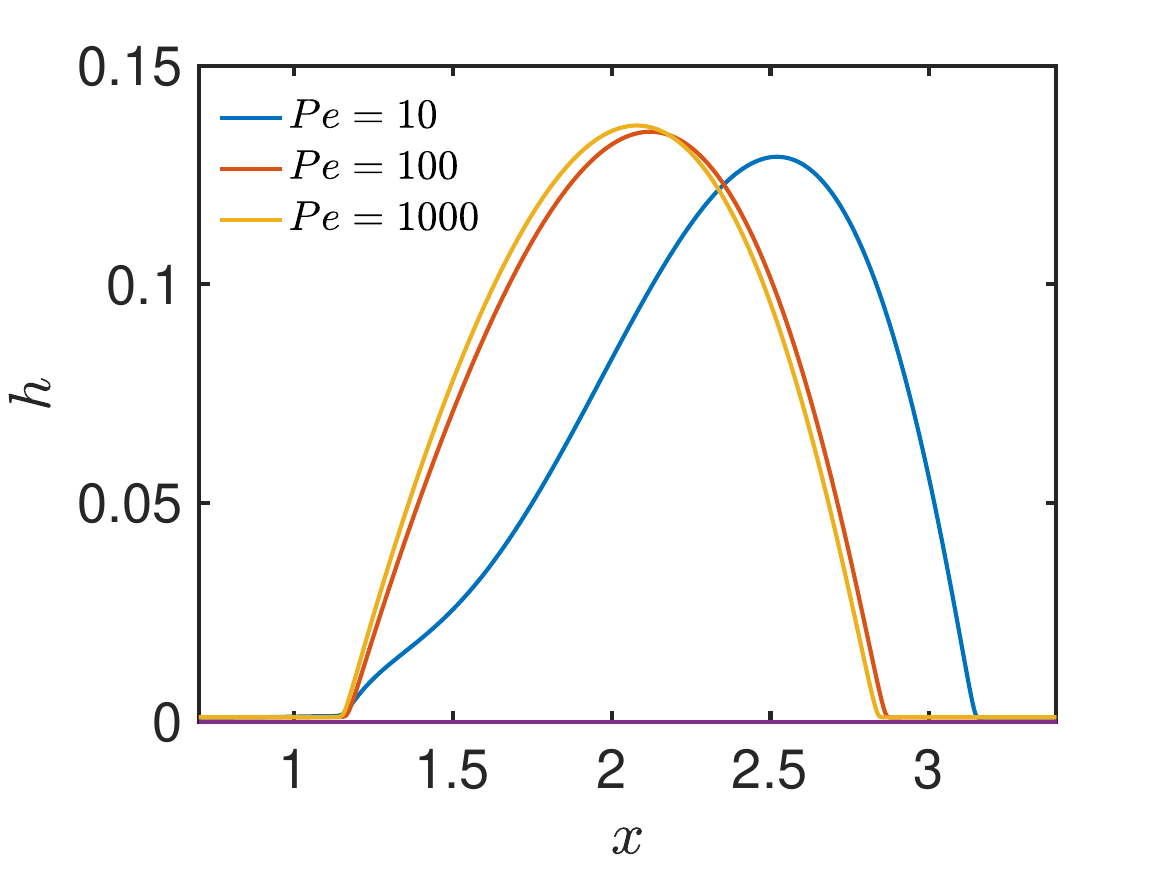}
\caption{}
\label{fig:droplet profiles on smooth for diff Pe}
\end{subfigure}
\begin{subfigure}{0.348\textwidth}
\includegraphics[width=\linewidth]{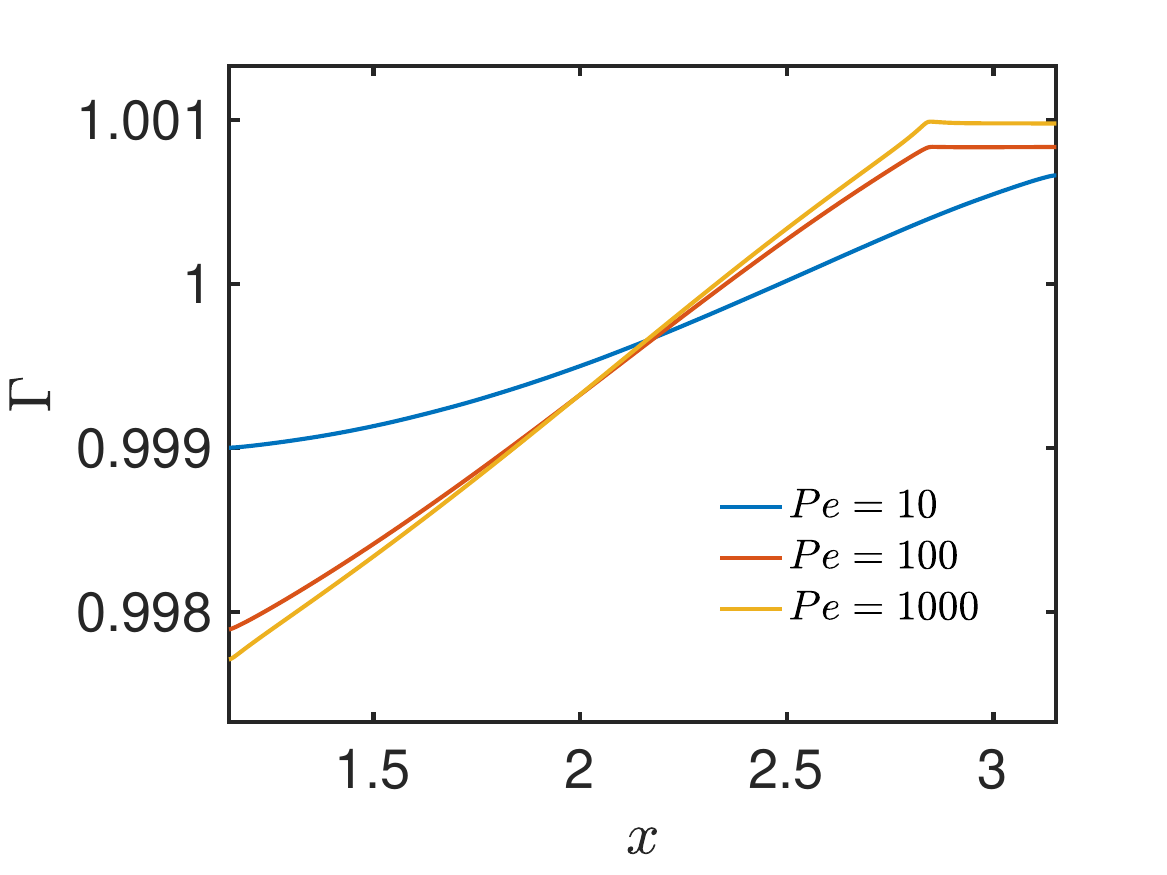}
\caption{}
\label{fig:surf conc on smooth for diff Pe}
\end{subfigure}
\begin{subfigure}{0.348\textwidth}
\includegraphics[width=\linewidth]{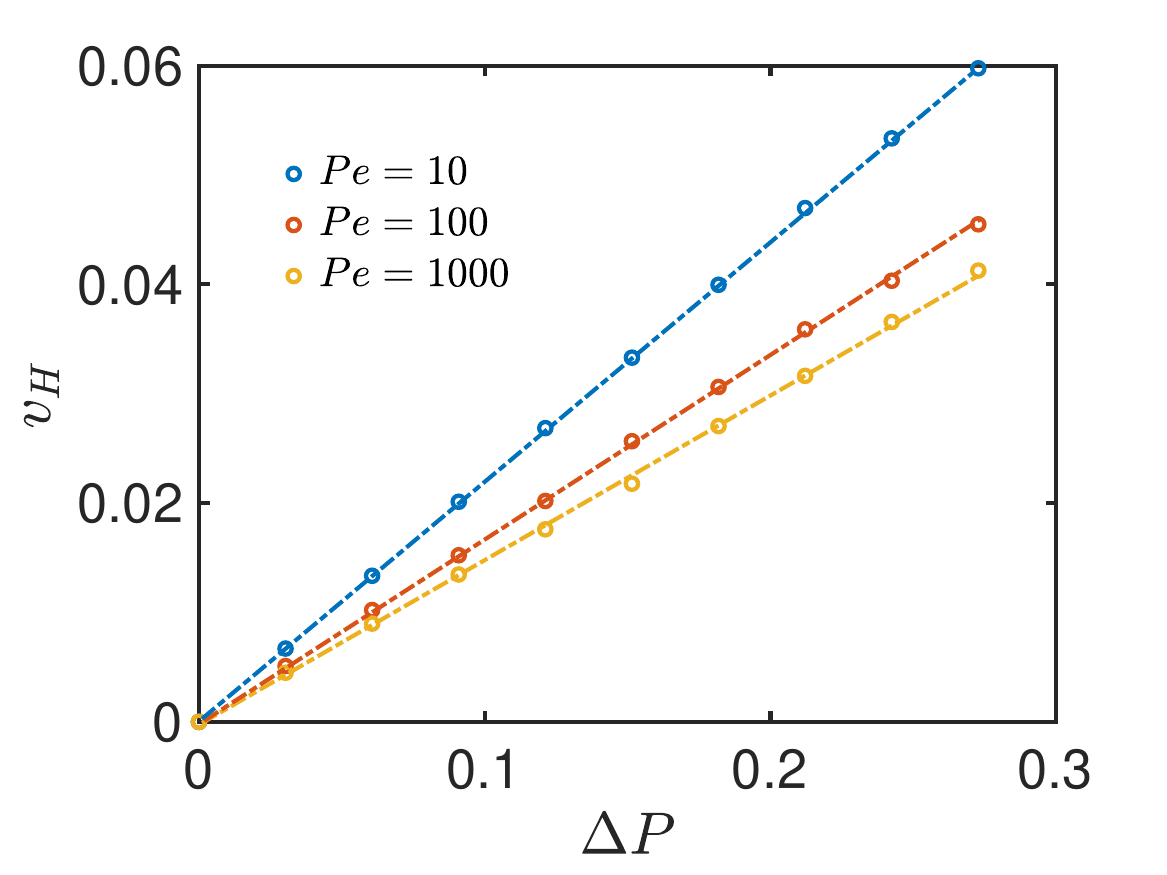}
\caption{}
\label{fig:vt analytical vs dP variation with Pe}
\end{subfigure}
\caption{(\subref{fig:vt vs dP variation with Pe}) $v_t$ vs. $\Delta P$ for different $Pe$. The open circles show numerical calculations and the dashed lines show linear fits. (\subref{fig:droplet profiles on smooth for diff Pe}) Steady droplet shapes for different $Pe$ with $\Delta P = 0.03$. The droplets have been shifted such that their receding contact lines coincide. (\subref{fig:surf conc on smooth for diff Pe}) Steady surfactant-concentration profiles ($\Gamma$ vs. $x$) for different $Pe$ with $\Delta P = 0.03$. (\subref{fig:vt analytical vs dP variation with Pe}) $v_H$ vs. $\Delta P$ for different $Pe$. The open circles show numerical calculations and the dashed lines show linear fits. The other parameters are $M = 100$, $L^*=6$, $\mu_r = 0.01$, $v_0 = 0.2$, $\theta_{eq}=10^{\circ}$ ($A=10^5$), and $b=0.001$ }
\label{fig:fig4}
\end{figure}

We consider droplet motion on a smooth substrate, which corresponds to $\eta(x)=0$, before investigating the influence of substrate topography on droplet dynamics. In our prior work \cite{mhatre2024shear}, we have shown that in the absence of external flow ($\Delta P = 0$), a perfectly wetting droplet ($\theta_{eq} = 0$) spreads in a manner which is consistent with Tanner's law for two-dimensional droplets ($x_{acl} - x_c \sim t^{1/7}$, where $x_c$ is the droplet center). A partially wetting droplet spreads until it reaches a steady shape with a contact angle within $1^{\circ}$ of the specified $\theta_{eq}$. 

We study the influence of flow of the surrounding fluid on droplet dynamics using the following methodology. First, to obtain an initial condition, a simulation is performed for a stationary surrounding fluid such that the droplet reaches a steady shape. Next, the surfactant concentration is initialized at the interface as $\Gamma(x,0)=1$, which corresponds to the concentration being equal to the mean concentration at all locations along the interface. Lastly, a negative dimensionless horizontal pressure gradient, $\Delta P$, is introduced in the channel. This methodology is motivated by the experiments discussed in \S\ref{sec:Intro} \cite{rowe2002oil, thoreau_physico-chemical_2006, freville2014effect}, where the droplet is allowed to reach a steady shape before introducing surfactant in the system.

Similar to the surfactant-free case\cite{mhatre2024shear}, for a given set of parameters, the droplet attains a steady shape and slides on the substrate with a terminal sliding velocity $v_t$, which is determined from the slope of the $x_{acl}$ vs. $t$ line. Figure \ref{fig:vt vs dP variation with M} shows the variation of $v_t$ with $\Delta P$ for three different values of $M$, where the open circles are results from numerical calculations, and the dashed lines are linear fits. It is seen that $v_t \sim \Delta P$ for all $M$ values, and $v_t$ decreases with $M$ for a given $\Delta P$. Figure \ref{fig:droplet profiles on smooth for diff M} shows the steady droplet shapes for different $M$ values, where the droplet is spread out and short for $M=1$, and becomes taller and narrower as $M$ increases. The steady surfactant-concentration profiles at the interface are shown in figure \ref{fig:surf conc on smooth for diff M}, and their temporal evolution is presented in appendix A1. The applied pressure gradient convects surfactant from left to right, creating a high-concentration region near the advancing contact line.   The resulting surface-tension gradient drives a Marangoni flow from the advancing to the receding contact line, leading to a taller droplet.  This Marangoni flow also convects surfactant toward the receding contact line, leading to a flatter steady-state concentration profile as $M$ increases. As the Marangoni flow opposes the direction of droplet motion, the droplet slides at a slower terminal velocity.

We also obtain an analytical solution using a flat-interface approximation to explain the above numerical solutions.  Similar to our prior work in the absence of surfactants \cite{mhatre2024shear}, we consider a steady pressure-driven two-layer flow having a flat interface, with the bottom layer being the droplet liquid and the top layer being the surrounding fluid.  We assume a constant surfactant concentration gradient, $(\partial \Gamma/\partial x)_m$, at the interface. We non-dimensionalize the physical quantities using the characteristic scales in \S\ref{subsec: governing equations} and obtain the following dimensionless velocity at the interface:
\begin{gather}
v_H = \frac{3H(1-H)(\Delta P - 2M(\partial\Gamma/\partial x)_{m})}{2 (1+H(\mu_r - 1))}, \label{eq: horizontal velocity flat-interface approximation}
\end{gather}
where $H$ is the dimensionless height of the flat interface. From the numerical calculations for given $\Delta P$, $M$, and $Pe$, we extract the maximum height of the steady droplet shape, $h_{max}$, and the average surfactant-concentration gradient along the interface, $(\partial \Gamma / \partial x) _{avg}$, by computing $\partial \Gamma / \partial x$ at all nodes along the interface and calculating their average value. We set $H = h_{max}$ and $(\partial \Gamma /\partial x)_m = (\partial \Gamma / \partial x)_{avg}$, calculate $v_H$ using (\ref{eq: horizontal velocity flat-interface approximation}), and compare it with $v_t$.

Figure \ref{fig:vt analytical vs dP variation with M} shows the variation of $v_H$ with $\Delta P$ for different $M$ values. It can be seen that the flat-interface approximation reproduces the qualitative trend observed in numerical simulations, as $v_H \sim \Delta P$, and $v_H$ decreases with $M$. However, it overpredicts the terminal sliding velocity by two orders of magnitude. This is likely because the flat-interface approximation does not account for viscous dissipation and surface-tension forces near the droplet contact line, which slow down droplet motion.

Next, we fix $M = 100$ and vary $Pe$. Figure \ref{fig:vt vs dP variation with Pe} shows the variation of $v_t$ with $\Delta P$ for three different values of $Pe$, where the open circles are results from numerical calculations, and the dashed lines are linear fits. It can be seen that $v_t \sim \Delta P$ for all $Pe$ values, and $v_t$ decreases with $Pe$. Figure \ref{fig:droplet profiles on smooth for diff Pe} shows the steady droplet shapes for different $Pe$ values, where the droplet is spread out and short for $Pe=10$, and becomes taller and narrower for larger $Pe$ values. The steady surfactant concentration profiles are shown in figure \ref{fig:surf conc on smooth for diff Pe}, and their temporal evolution is presented in appendix A1.  Increasing $Pe$ weakens surfactant diffusion, leading to larger surfactant concentration gradients and thus stronger Marangoni flows.  Thus, similar to figure \ref{fig:surf conc on smooth for diff M},  the steady-state concentration profiles become flatter as $Pe$ increases, and the droplet becomes taller and narrower.  As before, since the Marangoni flow opposes the direction of droplet motion, the droplet slides at a slower terminal velocity as $Pe$ increases. 

Figure \ref{fig:vt analytical vs dP variation with Pe} shows the variation of $v_H$ with $\Delta P$ for different $Pe$ values. Once again, it is seen that the flat-interface approximation predicts a linear relationship between the terminal sliding velocity and the applied pressure gradient ($v_H \sim \Delta P$),  and qualitatively captures the influence of $Pe$ on the terminal sliding velocity. But, it overpredicts the terminal sliding velocity by two orders of magnitude as it likely does not account for the viscous dissipation and surface-tension forces near the droplet contact line. Despite these shortcomings, the flat-interface approximation still provides useful insights even for rough substrates, as will be discussed in \S\ref{sec:surfactant influence}.

\section{Droplet dynamics on a rough substrate}
\label{sec:rough substrate}

\subsection{Pinning-depinning transition}
\label{subsec:pinning-depinning transition}

\begin{figure}[t]
\centering
\begin{subfigure}{0.4035\textwidth}
\includegraphics[width=\linewidth]{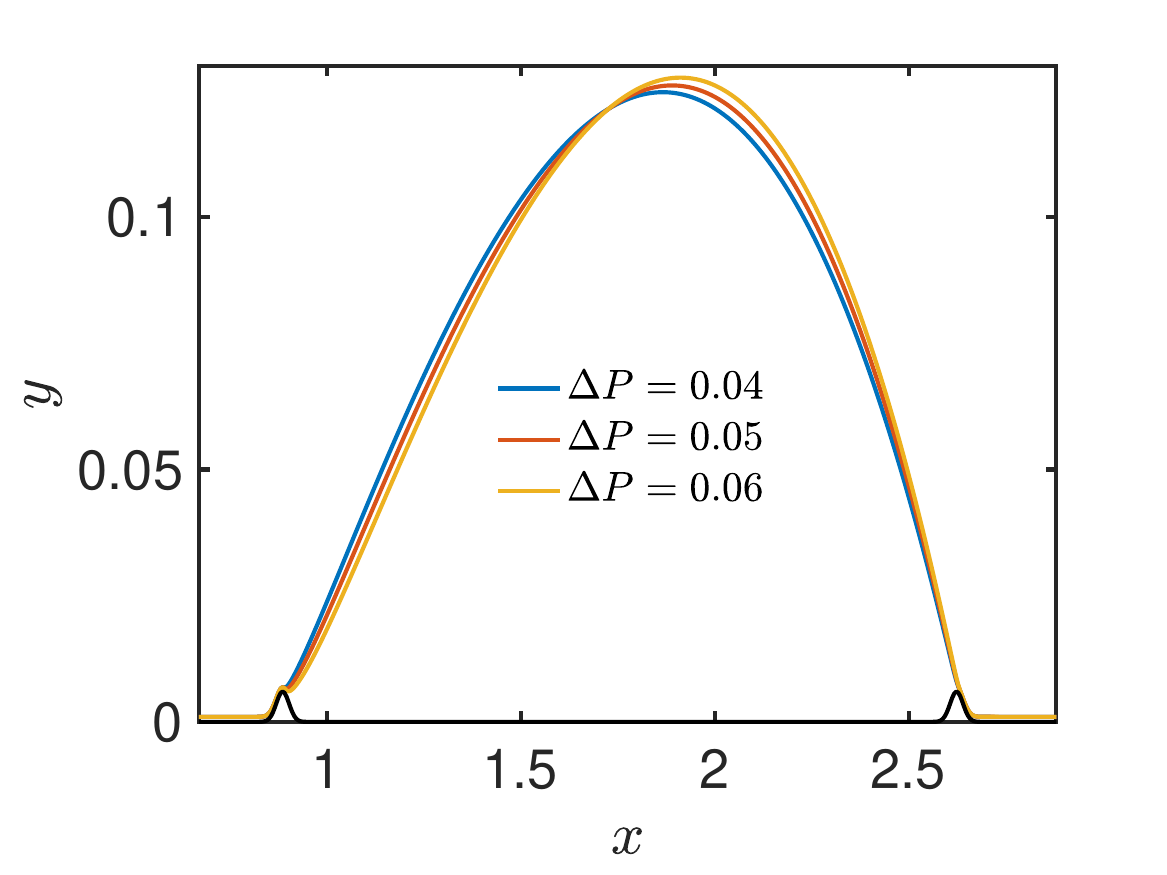}
\caption{}
\label{fig:steady droplet profiles for different dP}
\end{subfigure}
\begin{subfigure}{0.4035\textwidth}
\includegraphics[width=\linewidth]{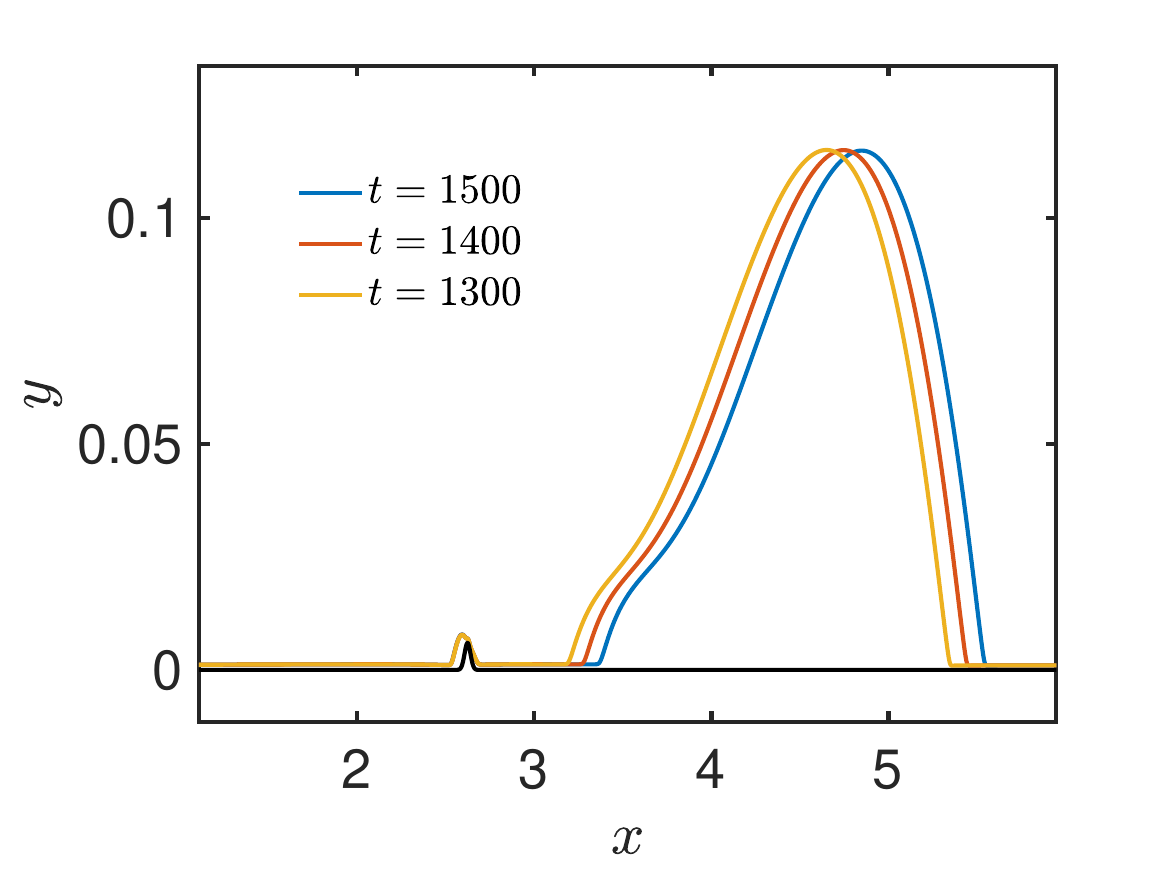}
\caption{}
\label{fig:transient motion of droplet for dP > dP_crit}
\end{subfigure}
\hfill
\begin{subfigure}{0.4035\textwidth}
\includegraphics[width=\linewidth]{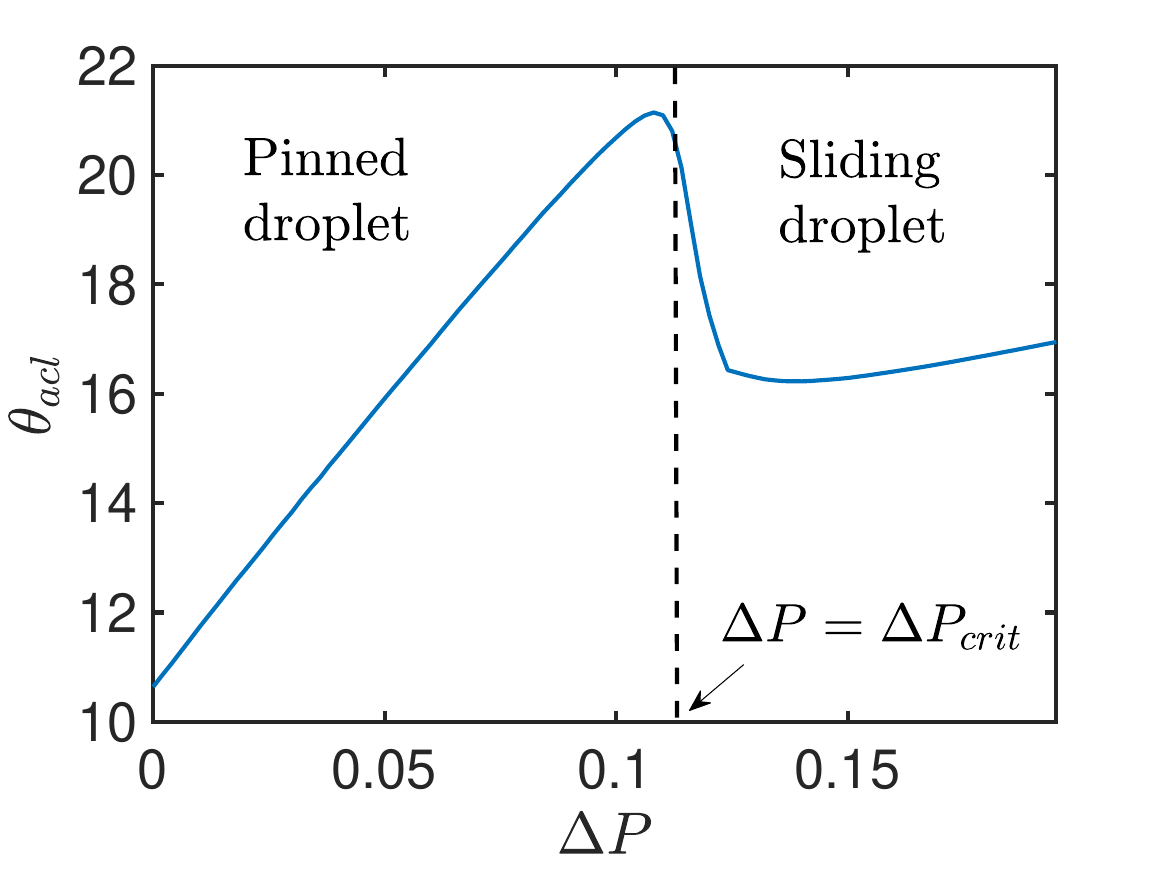}
\caption{}
\label{fig:sample solution family}
\end{subfigure}
\caption{(\subref{fig:steady droplet profiles for different dP}) Steady droplet shapes for different $\Delta P<\Delta P_{crit}$. (\subref{fig:transient motion of droplet for dP > dP_crit}) Droplet profiles at different $t$ for $\Delta P = 0.15$. The solid black lines show substrate topography and the other lines show droplet profiles. (\subref{fig:sample solution family}) Steady $\theta_{acl}$ vs. $\Delta P$. The other parameters are $M=1$, $Pe = 1000$, $L^*=6$, $\mu_r = 0.01$, $v_0 = 0.2$, $\theta_{eq}=10^{\circ}$ ($A=10^5$), $b=0.001$, $h_d = 0.05h_{max}$, and $w_d=2.5h_d$.}
\label{fig:fig5}
\end{figure}

We first briefly characterize the influence of substrate topography on droplet dynamics before considering in more detail the influence of surfactant in \S5.  An initial condition is obtained by performing a numerical simulation for a droplet without surfactant at the interface in a stationary fluid until the droplet reaches a steady shape. Next, the surfactant concentration is initialized at the interface as $\Gamma(x,0)=1$, which corresponds to the concentration being equal to the mean concentration at all locations along the interface. Then, a Gaussian-shaped bump-type defect is added at each contact line with $x_d=0$ (figure \ref{fig:schematic1}) to pin the droplet, such that $h_d = 0.05h_{max}$, and $w_d = 2.5h_d$, where $h_{max}$ is the maximum height of the steady droplet shape used as the initial condition. Lastly, a negative horizontal pressure gradient of magnitude $\Delta P$ is introduced in the channel. 

We follow the procedure used in our prior work \cite{mhatre2024shear} to determine the critical pressure gradient, $\Delta P_{crit}$, required for depinning of the droplet from the Gaussian-shaped bumps. This can also be interpreted as a  critical flow rate above which the droplet slides freely on the substrate in the experiments discussed in \S\ref{sec:Intro}. Below $\Delta P_{crit}$, the droplet remains pinned at the bumps and attains a deformed steady shape, and this deformation increases with $\Delta P$ as shown in figure \ref{fig:steady droplet profiles for different dP}. Consequently, the steady $\theta_{acl}$ attained by the droplet increases with $\Delta P$. Above $\Delta P_{crit}$, the droplet depins and slides on the substrate with a less deformed steady shape and a smaller $\theta_{acl}$ as shown in figure \ref{fig:transient motion of droplet for dP > dP_crit}. We plot the variation of the steady $\theta_{acl}$ attained by the droplet with $\Delta P$ in figure \ref{fig:sample solution family}, where $\Delta P_{crit}$ is identified as the $\Delta P$ at which there is a sharp decrease in $\theta_{acl}$.  The behavior observed in Figure \ref{fig:fig5} is similar to what is observed for surfactant-free droplets.\cite{mhatre2024shear}

\subsection{Depinning mechanism}
\label{subsec:depinning mechanism}

\begin{figure}[t]
\centering
\includegraphics[width = 0.61\linewidth]{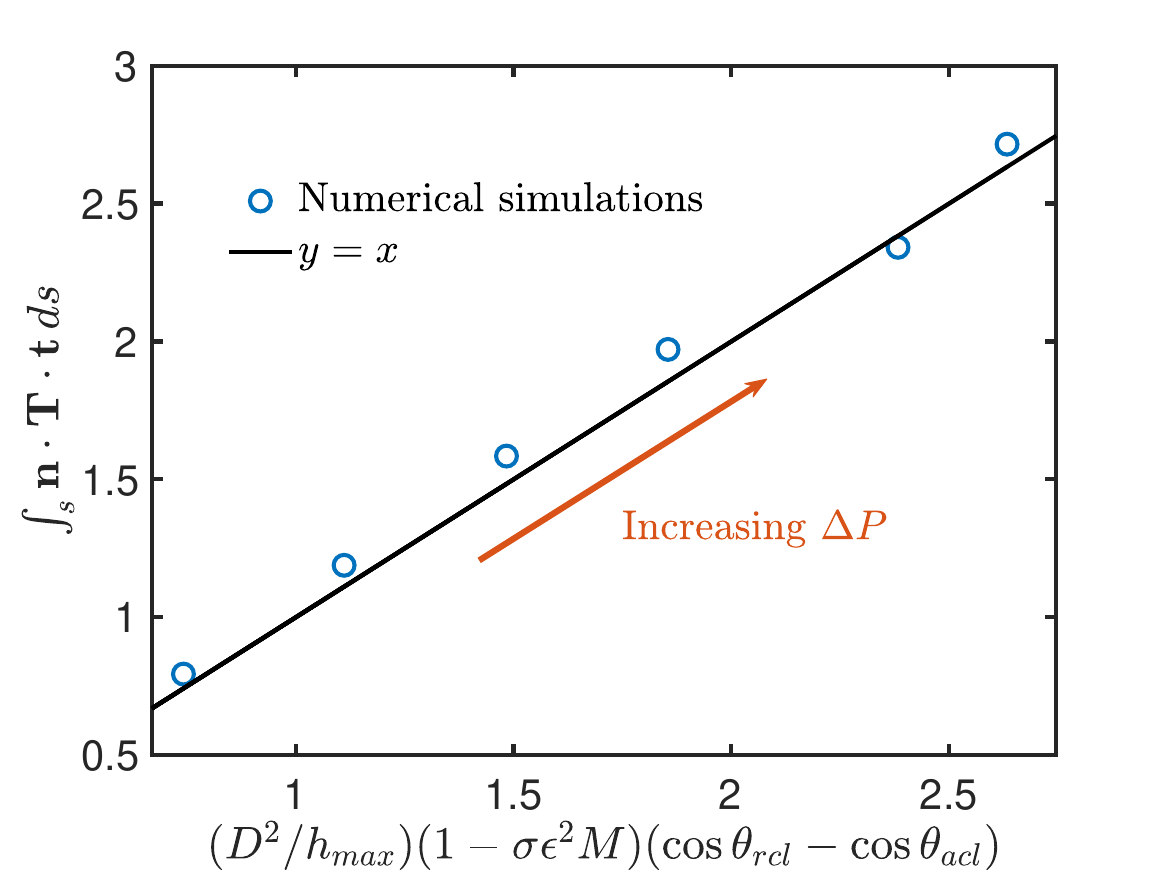}
\caption{Absolute values of the terms in the force-balance model for pinned droplets at different $\Delta P < \Delta P_{crit}$. The open blue circles are results from numerical simulations and the straight black line has a slope of unity.}
\label{fig:fig6}
\end{figure}

We use ideas from the force-balance model described in \S\ref{sec:Intro} to explain our numerical findings, where the drag force applied on the droplet due to the flow of surrounding fluid drives depinning, and the surface-tension force acting along the droplet contact line resists depinning. We have shown previously\cite{mhatre2024shear} that the drag force is nearly equal to the shear force for thin droplets, and the balance between the shear force and the surface-tension force in its dimensionless form can be written as $(D^2/h_{max})(\cos{\theta_{rcl}} - \cos{\theta_{acl}}) \sim \int_{s} \mathbf{n}\cdot\mathbf{T}\cdot\mathbf{t} \,ds$. In the presence of surfactant at the droplet interface, this force balance becomes, 
\begin{gather}
(D^2/h_{max})(1-\sigma \epsilon^2 M)(\cos{\theta_{rcl}} - \cos{\theta_{acl}}) \sim \int_{s} \mathbf{n}\cdot\mathbf{T}\cdot\mathbf{t} \,ds. \label{eq:Dimensionless force balance on smooth substrate}
\end{gather}
Here, $h_{max}$ is the maximum droplet height, $D = x_{acl} - x_{rcl}$ is the droplet width, $\mathbf{n}$ is the unit normal vector at the interface that points into the surrounding fluid, $\mathbf{T}$ is the droplet stress tensor, $\mathbf{t}$ is the unit tangent vector at the interface, and $s$ is the interface arclength coordinate such that $s = 0$ at the receding contact line and $s = 1$ at the advancing contact line.

The terms in the dimensionless force balance are calculated by extracting the values of $h_{max}$, $D$, $\theta_{acl}$, $\theta_{rcl}$, $\mathbf{n}$, $\mathbf{T}$, and $\mathbf{t}$ from the steady droplet shapes obtained from numerical simulations. Figure \ref{fig:fig6} shows these terms for different values of $\Delta P < \Delta P_{crit}$, where it is seen that there is a linear relationship between $(D^2/h_{max})(1-\sigma \epsilon^2 M)(\cos{\theta_{rcl}} - \cos{\theta_{acl}})$ and $\int_{s} \mathbf{n}\cdot\mathbf{T}\cdot\mathbf{t} \,ds$, which is consistent with the dimensionless force balance. Above $\Delta P_{crit}$, $\int_{s} \mathbf{n}\cdot\mathbf{T}\cdot\mathbf{t} \,ds$ exceeds $(D^2/h_{max})(1-\sigma \epsilon^2 M)(\cos{\theta_{rcl}} - \cos{\theta_{acl}})$, leading to droplet depinning. Similar to the case of surfactant-free droplets \cite{mhatre2024shear}, the pinning locations of the receding and advancing contact lines, $x_{rcl}$ and $x_{acl}$, coincide with the points on the defects where there is a maximum negative slope, as this maximizes the surface-tension force $(D^2/h_{max})(1-\sigma \epsilon^2 M)(\cos{\theta_{rcl}} - \cos{\theta_{acl}})$. Note that surfactants enter the force balance in two ways.  First, they affect the shear force via their influence on $\mathbf{T}$ and the droplet shape.  Second, they influence the mean interfacial tension via the $\sigma \epsilon^2 M$ term.  

\section{Influence of Marangoni number and Peclet number on droplet dynamics}
\label{sec:surfactant influence}

\begin{figure}[t]
\centering
\includegraphics[width = 0.61\linewidth]{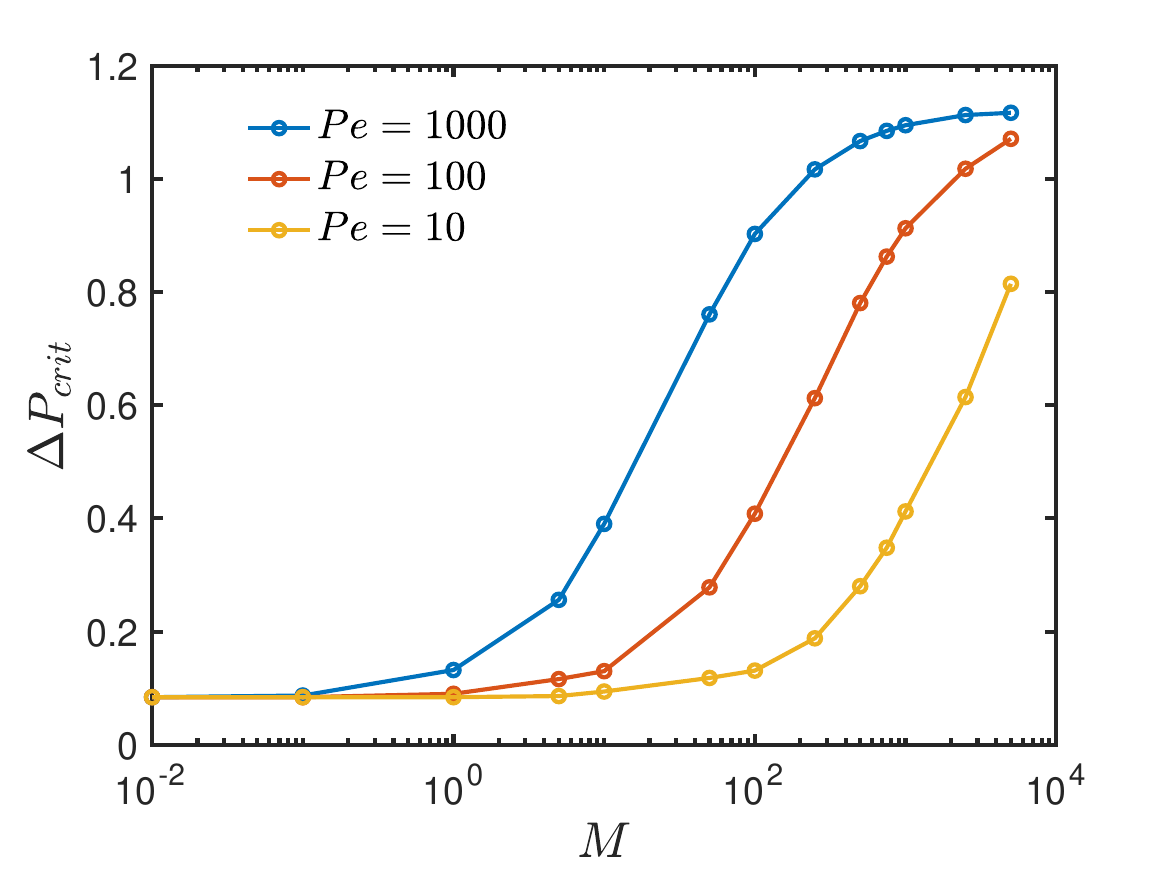}
\caption{$\Delta P_{crit}$ vs. $M$ for different $Pe$ values. The other parameters are $L^*=6$, $\mu_r = 0.01$, $v_0 = 0.2$, $\theta_{eq}=10^{\circ}$ ($A=10^5$), $b=0.001$, $h_d = 0.05h_{max}$, and $w_d=2.5h_d$.}
\label{fig:fig7}
\end{figure}

\begin{figure}[t]
\centering
\begin{subfigure}{0.4035\textwidth}
\includegraphics[width=\linewidth]{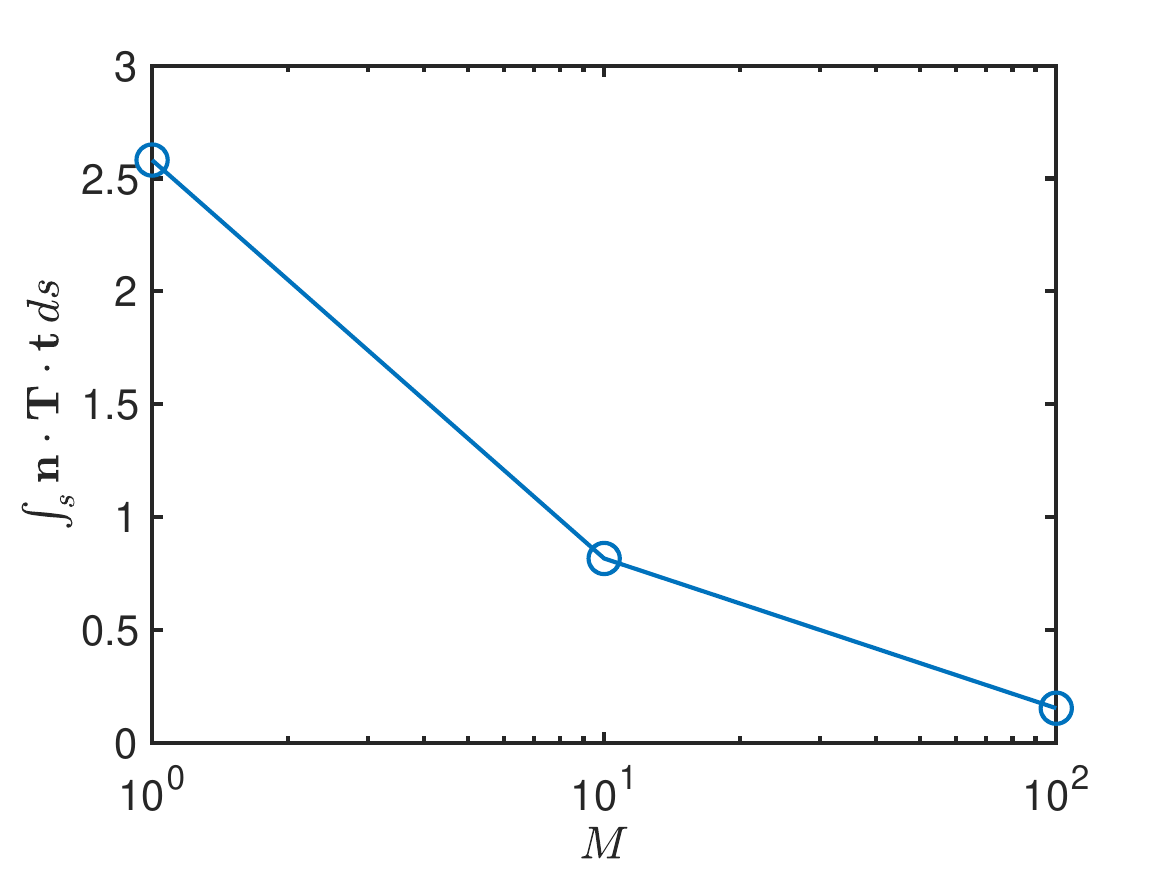}
\caption{}
\label{fig:shear force vs M for rough substrate}
\end{subfigure}
\begin{subfigure}{0.4035\textwidth}
\includegraphics[width=\linewidth]{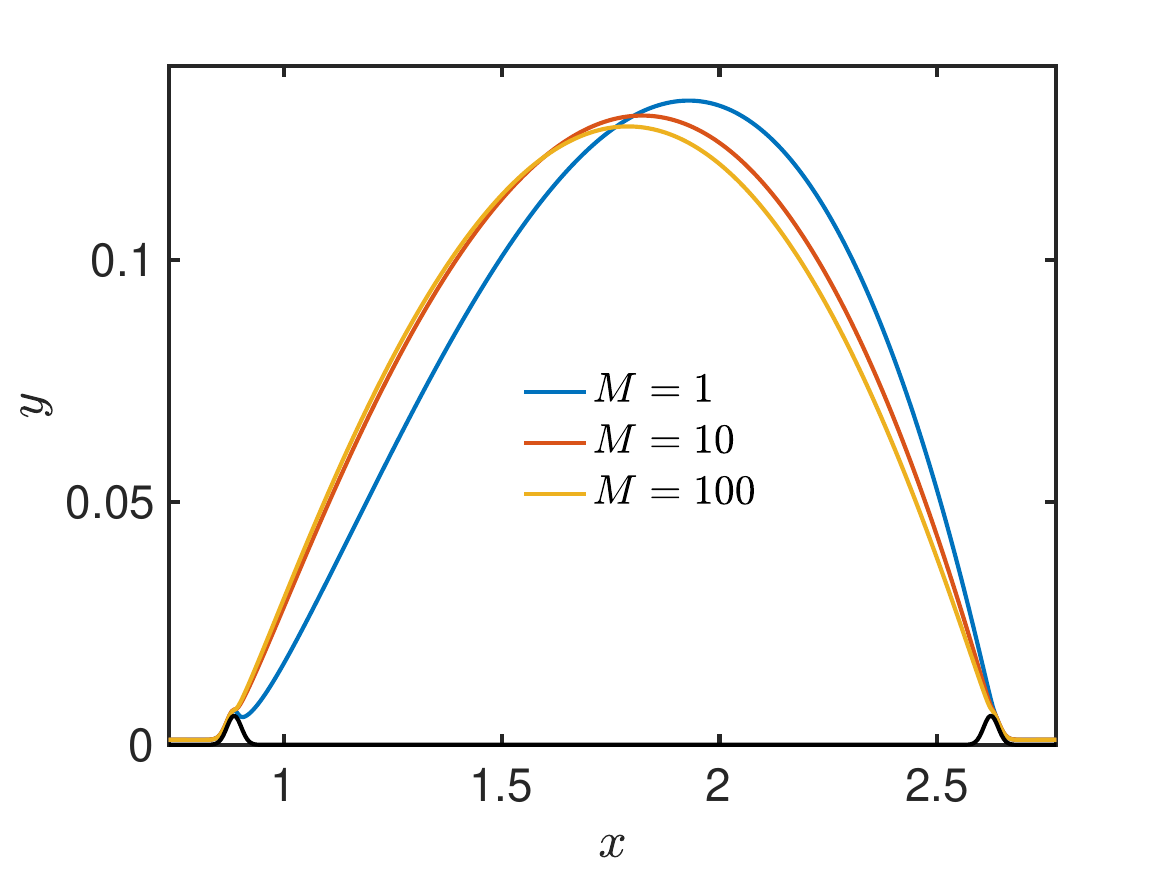}
\caption{}
\label{fig:steady droplet shapes vs M for rough substrate}
\end{subfigure}
\hfill
\begin{subfigure}{0.4035\textwidth}
\includegraphics[width=\linewidth]{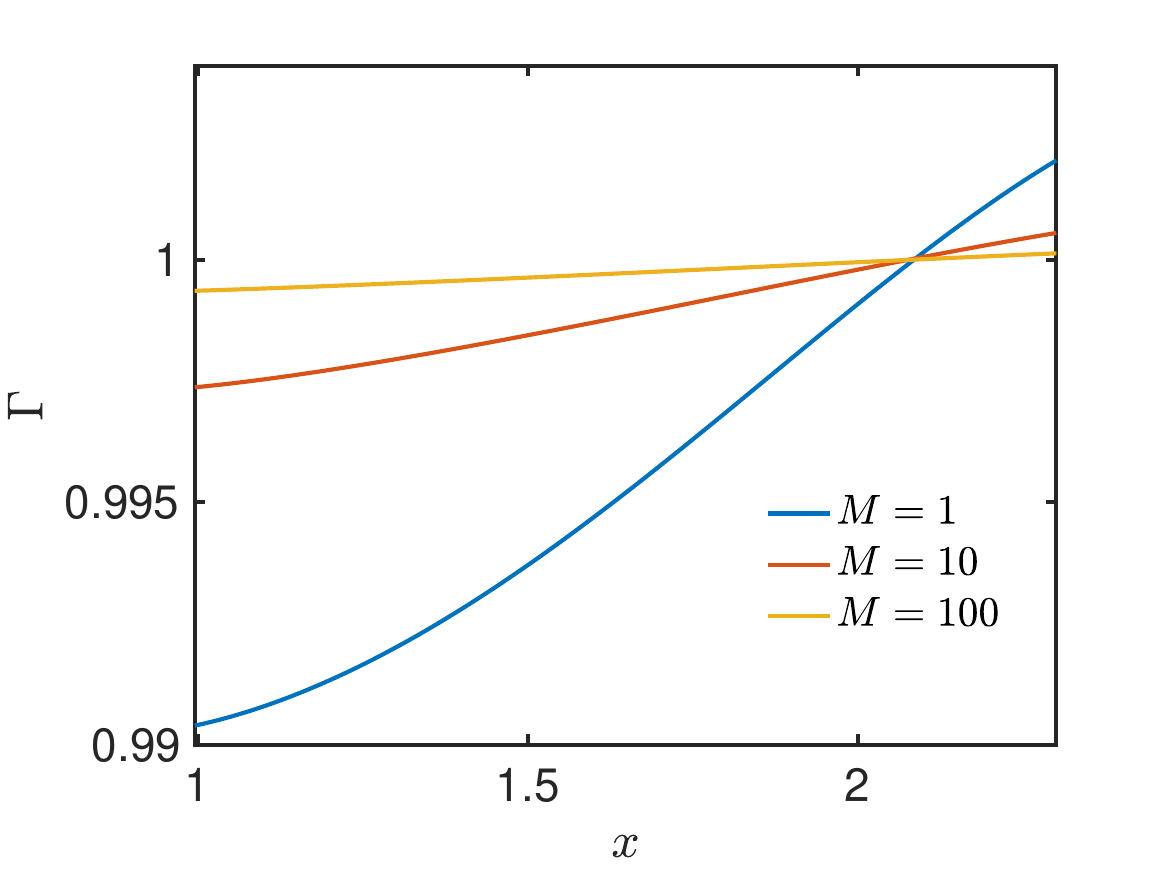}
\caption{}
\label{fig:surf conc profile vs M for rough substrate}
\end{subfigure}
\caption{(\subref{fig:shear force vs M for rough substrate}) Shear force exerted on the droplet vs. $M$. (\subref{fig:steady droplet shapes vs M for rough substrate}) Steady droplet profiles for different $M$. (\subref{fig:surf conc profile vs M for rough substrate}) Steady surfactant-concentration profiles for different $M$. The other parameters are $Pe=1000$, $\Delta P=0.07$, $L^*=6$, $\mu_r = 0.01$, $v_0 = 0.2$, $\theta_{eq}=10^{\circ}$ ($A=10^5$), $b=0.001$, $h_d = 0.05h_{max}$, and $w_d=2.5h_d$. Here, $\Delta P < \Delta P_{crit} = 0.1$.}
\label{fig:fig8}
\end{figure}

\begin{figure}[t]
\centering
\begin{subfigure}{0.4035\textwidth}
\includegraphics[width=\linewidth]{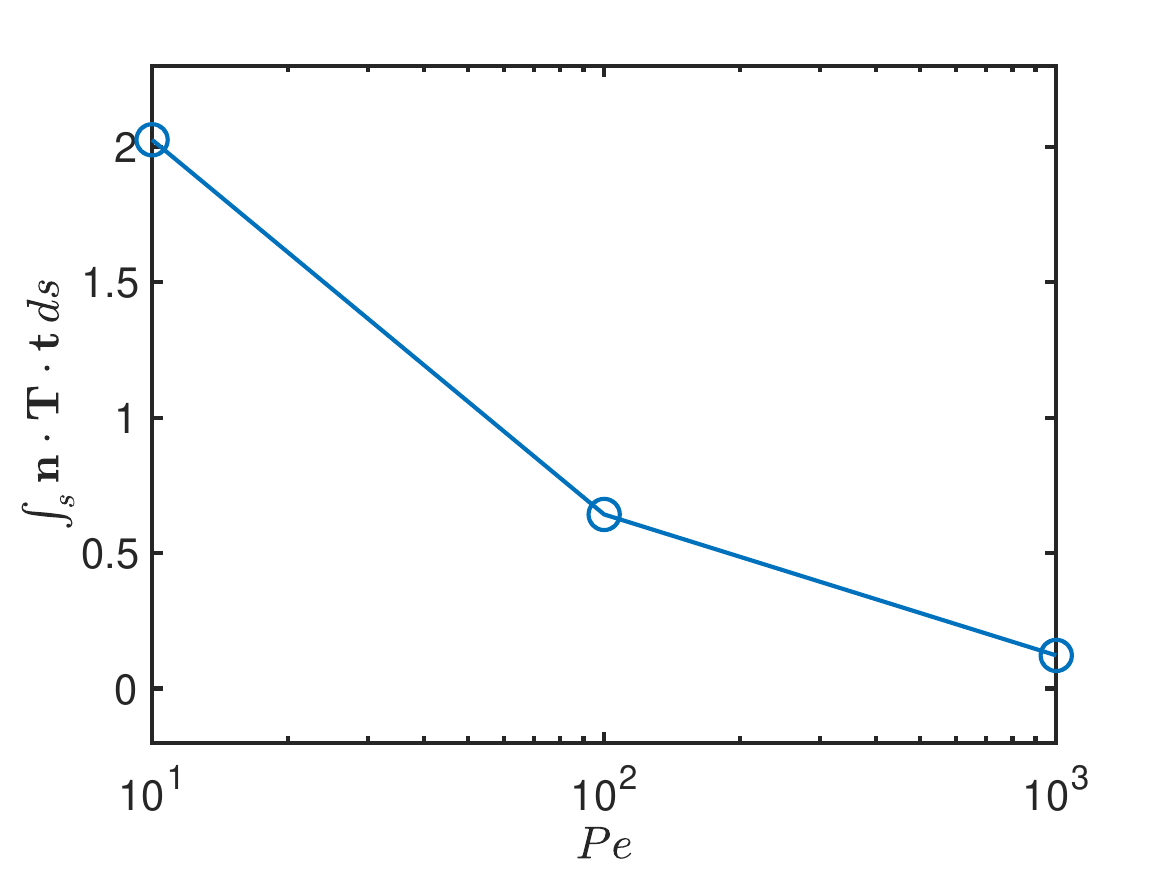}
\caption{}
\label{fig:shear force vs Pe for rough substrate}
\end{subfigure}
\begin{subfigure}{0.4035\textwidth}
\includegraphics[width=\linewidth]{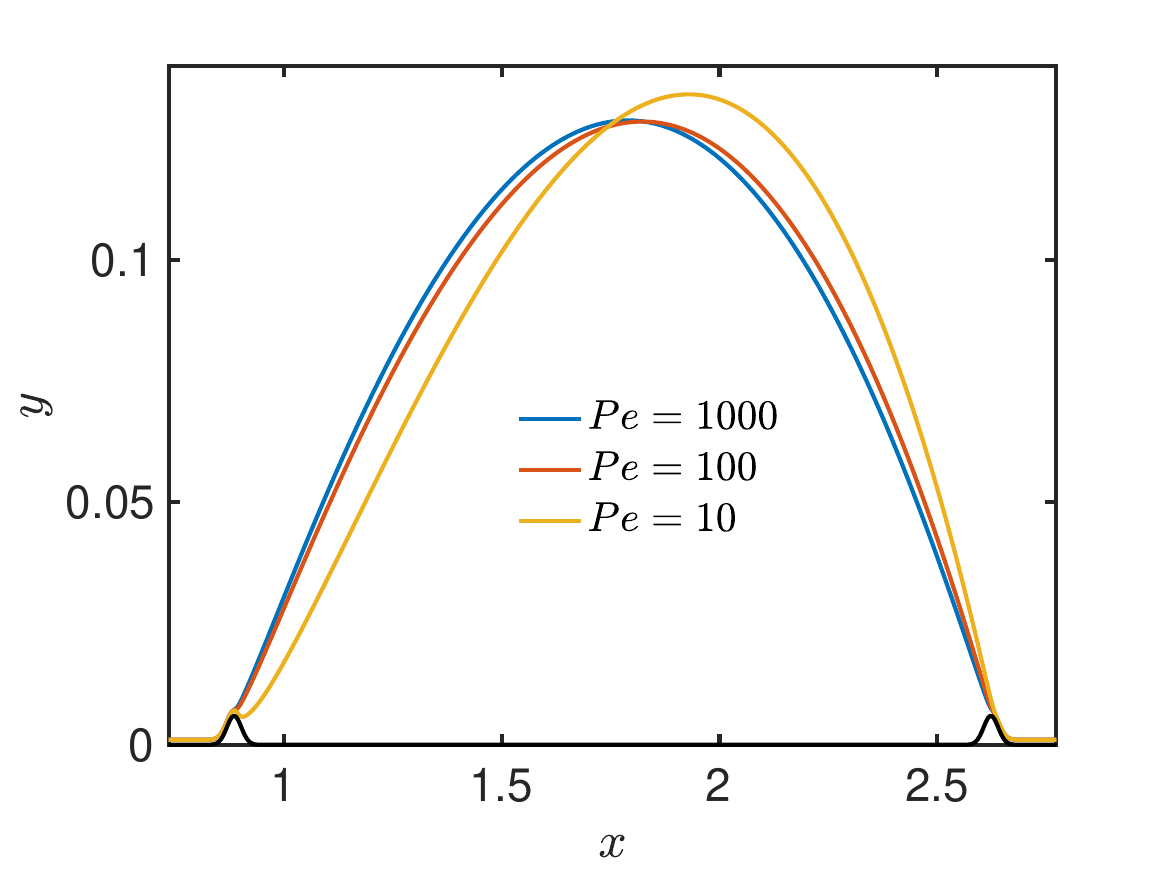}
\caption{}
\label{fig:steady droplet shapes vs Pe for rough substrate}
\end{subfigure}
\hfill
\begin{subfigure}{0.4035\textwidth}
\includegraphics[width=\linewidth]{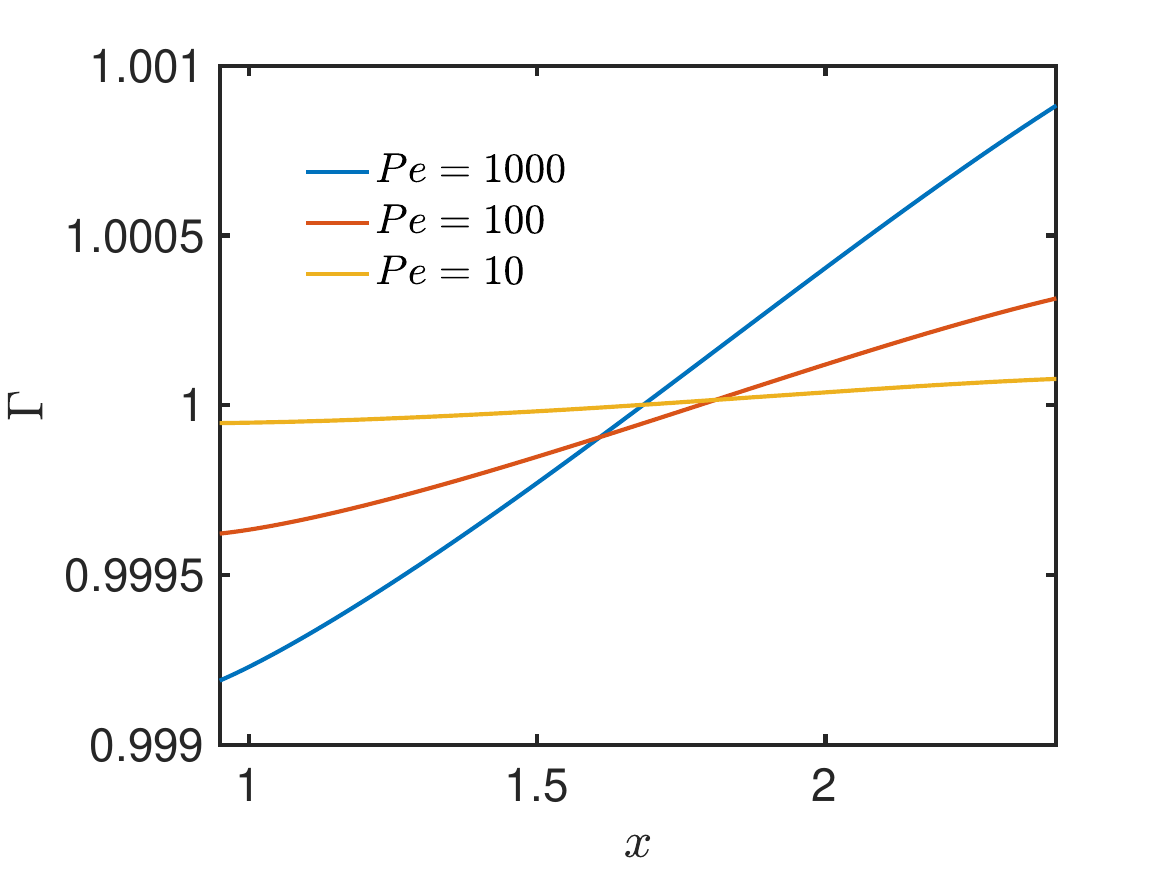}
\caption{}
\label{fig:surf conc profile vs Pe for rough substrate}
\end{subfigure}
\caption{(\subref{fig:shear force vs Pe for rough substrate}) Shear force exerted on the droplet vs. $Pe$. (\subref{fig:steady droplet shapes vs Pe for rough substrate}) Steady droplet profiles for different $Pe$. (\subref{fig:surf conc profile vs Pe for rough substrate}) Steady surfactant-concentration profiles for different $Pe$. The other parameters are $M = 100$, $\Delta P=0.07$, $L^*=6$, $\mu_r = 0.01$, $v_0 = 0.2$, $\theta_{eq}=10^{\circ}$ ($A=10^5$), $b=0.001$, $h_d = 0.05h_{max}$, and $w_d=2.5h_d$. Here, $\Delta P < \Delta P_{crit} = 0.08$.}
\label{fig:fig9}
\end{figure}

\begin{figure}[t]
\centering
\begin{subfigure}{0.49643\textwidth}
\includegraphics[width=\linewidth]{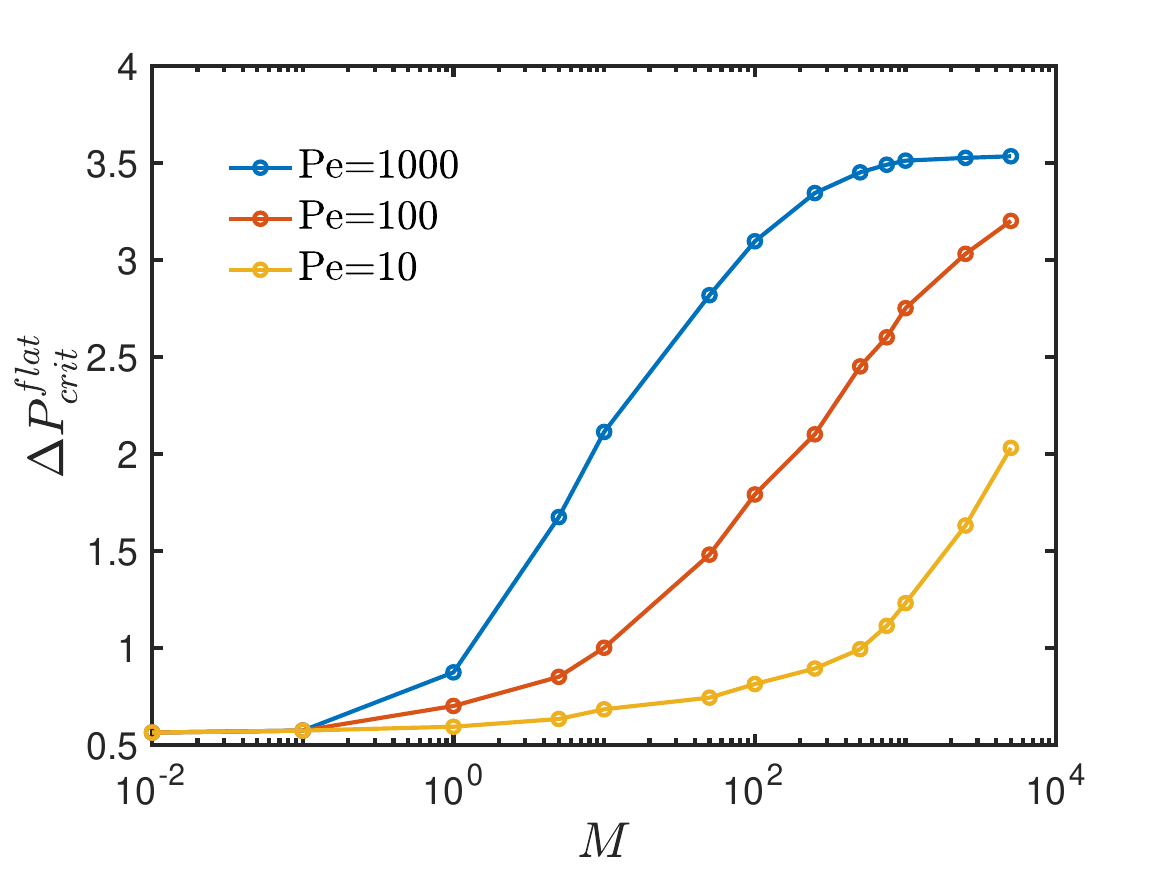}
\caption{}
\label{fig:dP crit vs M for all Pe analytical}
\end{subfigure}
\hfill
\begin{subfigure}{0.49643\textwidth}
\includegraphics[width=\linewidth]{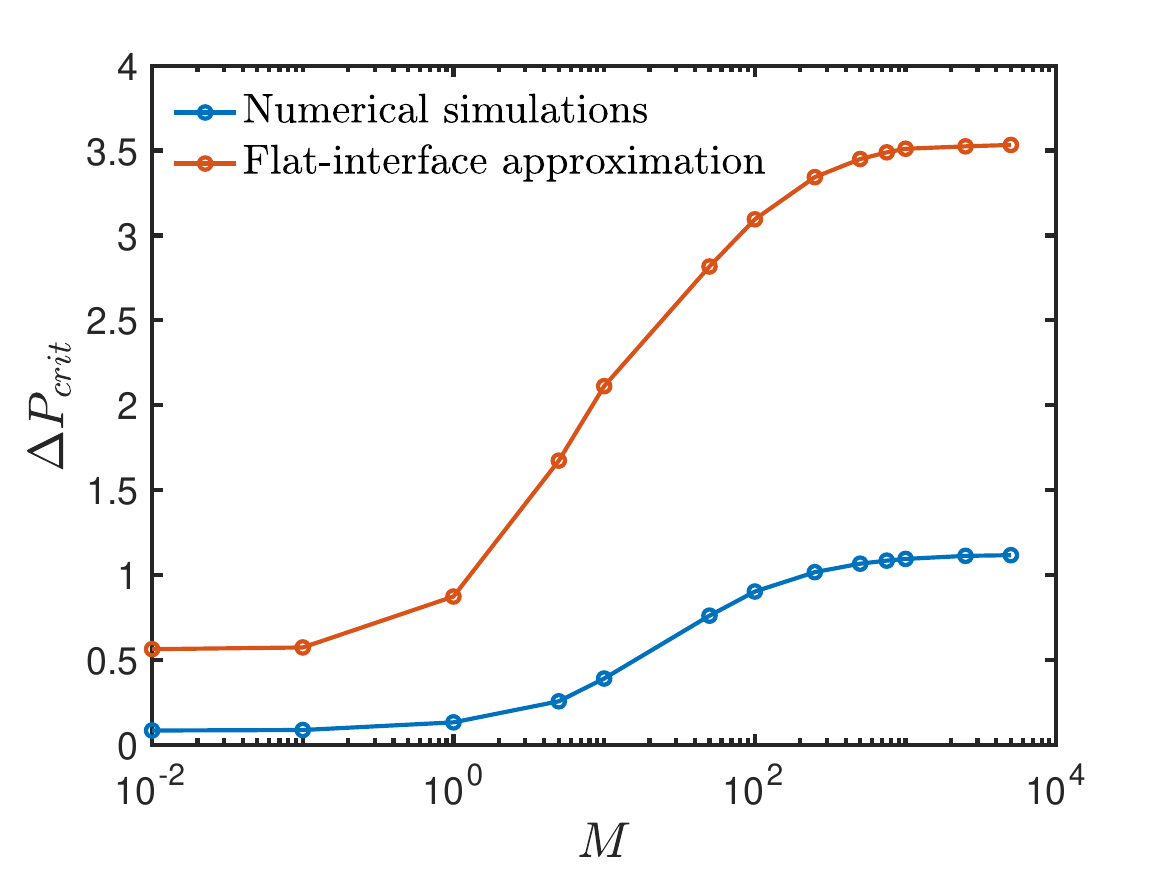}
\caption{}
\label{fig:dP crit vs M for Pe=1000}
\end{subfigure}
\caption{(\subref{fig:dP crit vs M for all Pe analytical}) $\Delta P_{crit}$ obtained using flat-interface approximation vs. $M$ for three different $Pe$ values. (\subref{fig:dP crit vs M for Pe=1000}) Comparison of numerical simulations with the flat-interface approximation for $Pe=1000$. The other parameters are $L^*=6$, $\mu_r = 0.01$, $v_0 = 0.2$, $\theta_{eq}=10^{\circ}$ ($A=10^5$), $b=0.001$, $h_d = 0.05h_{max}$, and $w_d=2.5h_d$.}
\label{fig:fig10}
\end{figure}

Figure \ref{fig:fig7} shows the variation of $\Delta P_{\mathrm{crit}}$ with $M$ for different $Pe$ values.  It is seen that $\Delta P_{\mathrm{crit}}$ increases as $M$ increases or as $Pe$ increases. The increase with $M$ is consistent with the shear-force ($\int_{s} \mathbf{n}\cdot\mathbf{T}\cdot\mathbf{t} \,ds$) calculations shown in figure \ref{fig:shear force vs M for rough substrate}, where the droplet experiences a lower shear force as $M$ increases for a fixed $\Delta P$. These findings can be rationalized by examining the steady shapes of pinned droplets for different $M$ values, as shown in figure \ref{fig:steady droplet shapes vs M for rough substrate}, where droplet deformation decreases with increasing $M$. Figure \ref{fig:surf conc profile vs M for rough substrate} shows the steady surfactant concentration profiles for different $M$ values. The applied pressure gradient drives a flow that transports surfactant from left to right, creating a high-concentration region near the advancing contact line.   The resulting Marangoni stresses drive surfactant in the opposite direction, leading to flatter steady-state concentration profiles and less deformed droplets as $M$ increases.  As the Marangoni stresses oppose the applied shear, a larger $\Delta P_{\mathrm{crit}}$ is required for droplet depinning.

Figure \ref{fig:fig7} also shows that $\Delta P_{\mathrm{crit}}$ increases as $Pe$ increases. This is consistent with the shear-force calculations shown in figure \ref{fig:shear force vs Pe for rough substrate}, where the shear force exerted on the droplet decreases as $Pe$ increases for a fixed $\Delta P$. This can be rationalized by examining the steady shapes of pinned droplets for different $Pe$ values, as shown in figure \ref{fig:steady droplet shapes vs Pe for rough substrate}, where droplet deformation decreases with increasing $Pe$. Figure \ref{fig:surf conc profile vs Pe for rough substrate} shows the steady surfactant concentration profiles for different $Pe$ values. The applied pressure gradient drives flow of surfactant from left to right, creating a high-concentration region near the advancing contact line. The resulting Marangoni stresses drive surfactant in the opposite direction, leading to flatter steady-state concentration profiles and less deformed droplets as $Pe$ increases.  As the Marangoni stresses oppose the applied shear, a larger $\Delta P_{\mathrm{crit}}$ is required for droplet depinning.

We also obtain an analytical solution to estimate the critical pressure gradient. We calculate the shear force at the droplet interface, which reduces to $\int_{s} \partial u^d / \partial y \,ds$ within the limits of lubrication theory.  We apply the flat-interface approximation as discussed in \S\ref{sec:smooth substrate}, and substitute the flat-interface height, $H$, with the maximum height of the steady droplet shape, $h_{\mathrm{max}}$. 
The shear force is equated to the surface-tension force, $(D^2/h_{\mathrm{max}})(1-\sigma \epsilon^2 M)(\cos{\theta_{\mathrm{rcl}}} - \cos{\theta_{\mathrm{acl}}})$, to obtain an expression for the critical pressure gradient:
\begin{gather}
\Delta P_{\mathrm{crit}}^{\mathrm{flat}} = \frac{(D^2/h_{\mathrm{max}})(1-\sigma \epsilon^2 M)(\cos{\theta_{\mathrm{rcl}}} - \cos{\theta_{\mathrm{acl}}}) + 3\frac{1-h_{\mathrm{max}}}{1+h_{\mathrm{max}}(\mu_r-1)} M \partial \Gamma/\partial x}{(3h_{\mathrm{max}}^2\mu_r - 3(h_{\mathrm{max}}-1)^2)/(2(1+h_{\mathrm{max}}(\mu_r-1)))}. \label{eq: dP critical analytical} 
\end{gather}
The values of $D$, $h_{\mathrm{max}}$, $\theta_{\mathrm{rcl}}$, and $\theta_{\mathrm{acl}}$ are extracted from the steady droplet shapes obtained from numerical simulations just below $\Delta P_{\mathrm{crit}}$, 
and $\sigma$ and $\partial \Gamma/\partial x$ are calculated by taking an average of these quantities along the nodes located at the droplet interface. 

Figure \ref{fig:dP crit vs M for all Pe analytical} shows the variation of $\Delta P_{\mathrm{crit}}^{\mathrm{flat}}$ with $M$ for different $Pe$ values.  The critical pressure gradient increases as $M$ increases or as $Pe$ increases. This trend is qualitatively consistent with results from numerical simulations shown in figure \ref{fig:fig7}. Figure \ref{fig:dP crit vs M for Pe=1000} shows a comparison between $\Delta P_{\mathrm{crit}}$ predictions using the flat-interface approximation and numerical simulations for $Pe = 1000$, where both predictions are of the same order of magnitude, but the flat-interface approximation is off by a factor of about $4$. This discrepancy is likely due to several assumptions used to derive the shear force, such as the interface being flat and the surfactant concentration gradient being constant. Although the flat-interface approximation does not quantitatively match the numerical results, it qualitatively captures the influence of $M$ and $Pe$ on $\Delta P_{\mathrm{crit}}$. 

The findings presented in this section show that the presence of insoluble surfactant at the droplet interface increases the critical pressure gradient required for depinning. This trend contrasts with the experimental observations discussed in \S\ref{sec:Intro} \cite{thoreau_physico-chemical_2006}, where increasing the surfactant concentration decreased the critical flow rate for depinning. This difference likely arises because, in those experiments, the surfactant was introduced into the system through the surrounding fluid and adsorbed onto the droplet–fluid interface. Any surfactant concentration gradients that might arise during flow would likely be suppressed if the adsorption timescale were shorter than that of the flow. In such a case, Marangoni stresses would be negligible, and the uniform interfacial coverage would simply lower the surface-tension force along the contact line without reducing the shear force on the droplet, leading to depinning at a lower critical flow rate. In this work, the reduction in surface-tension force due to surfactant is accounted for through the term $\epsilon^2 \sigma M$ in (\ref{eq: normal stress balance}). However, neglecting this effect does not change the qualitative nature of the results (see appendix A2), which indicates that depinning is governed by Marangoni stresses rather than the overall reduction in interfacial tension, for droplets laden with insoluble surfactant.    

Incorporating adsorption–desorption kinetics and bulk surfactant transport into the modeling framework would help bridge the gap between experimental conditions described above\cite{thoreau_physico-chemical_2006} and the insoluble-surfactant regime examined here. The present framework is particularly relevant for applications such as droplet microfluidics \cite{sinz2012self, kovalchuk2023review}, where the surfactant is not soluble in the surrounding phase (typically air), so surfactant-induced Marangoni stresses can potentially be exploited to control droplet motion.

\section{Conclusions}
\label{sec:conclusions}

In this work, we have built upon the lubrication-theory-based framework introduced in our earlier study \cite{mhatre2024shear} to investigate the dynamics of surfactant-laden droplets on rough substrates subject to pressure-driven flow of an immiscible surrounding fluid, in the regime where the surfactant is confined to the interface. The model offers several advantages over force-balance approaches \cite{basu1997model}, as it resolves the transient evolution of the droplet shape, captures contact-line motion over prescribed surface topography, and accounts for surfactant transport along the interface and the resulting Marangoni stresses, predicting how they modify the critical pressure gradient required for droplet depinning. Simplified analytical models are also developed that successfully capture the qualitative trends observed in the numerical simulations.

Below a critical pressure gradient, $\Delta P_{crit}$, the shear force exerted on the droplet by the surrounding fluid is balanced by the surface-tension forces along the contact line, and both the receding and advancing contact lines remain pinned at the defects. Above $\Delta P_{crit}$, the shear force exceeds the surface-tension force, leading to depinning of the contact lines and subsequent sliding of the droplet along the substrate. The applied pressure gradient drives interfacial surfactant transport from the receding to the advancing side, resulting in accumulation of surfactant near the advancing contact line. This generates a Marangoni flow directed from the advancing to the receding side, which opposes the applied shear and decreases the net force acting to depin the droplet. As $M$ increases, the sensitivity of surface tension to the surfactant concentration becomes stronger, which strengthens the Marangoni flow and increases $\Delta P_{crit}$. As $Pe$ increases, convection dominates over diffusion and the surfactant-concentration profile is no longer flattened by interfacial diffusion, which also strengthens the Marangoni flow and increases $\Delta P_{crit}$.

The results of this study show that Marangoni stresses arising from surfactant-concentration gradients can significantly alter droplet dynamics by increasing the critical pressure gradient required for depinning. This trend contrasts with systems in which Marangoni effects may be negligible \cite{thoreau_physico-chemical_2006}, where the addition of surfactants typically lowers the critical flow rate for depinning, likely due to a reduction in the surface-tension force at the contact line without a significant change in the net depinning force. The ability to tune Marangoni stresses by controlling surfactant properties and substrate topography provides a potential mechanism to actively control droplet motion in applications such as droplet-based microfluidics, enhanced oil recovery, and surface cleaning. The modeling framework can be readily extended to investigate more complex scenarios, including chemically heterogeneous substrates through spatially varying disjoining pressures \cite{schwartz1998hysteretic, schwartz1998simulation}, three-dimensional substrate defects \cite{mhatre_pinningdepinning_2024}, and systems with soluble surfactants \cite{yang2024rupture}. In addition, thermal Marangoni effects \cite{hou2013temperature, mhatre2022delaying}, where temperature gradients are deliberately applied to manipulate interfacial tension, offer another promising route for controlling droplet behavior, and exploring these coupled physicochemical effects forms a compelling direction for future theoretical and experimental studies.

\section*{Acknowledgments} \label{sec:Acknowledgment}

This material is based upon work supported by the National Science Foundation under Grant No. CBET-1935968. 

\section*{Declaration of interests} \label{sec:declaration of interests}

The authors report no conflicts of interest.

\appendix
\section*{Appendix} \label{sec:appendix}
\renewcommand{\thefigure}{A\arabic{figure}}
\setcounter{figure}{0}  
\renewcommand{\thesubsection}{A\arabic{subsection}}
\setcounter{subsection}{0}  
\begin{appendices} \label{Appendix}
\makeatletter\def\@currentlabel{appendix}\makeatother

\subsection{Temporal evolution of surfactant concentration gradient for smooth substrates}
\begin{figure}[ht]
\centering
\begin{subfigure}{0.378\textwidth}
\includegraphics[width=\linewidth]{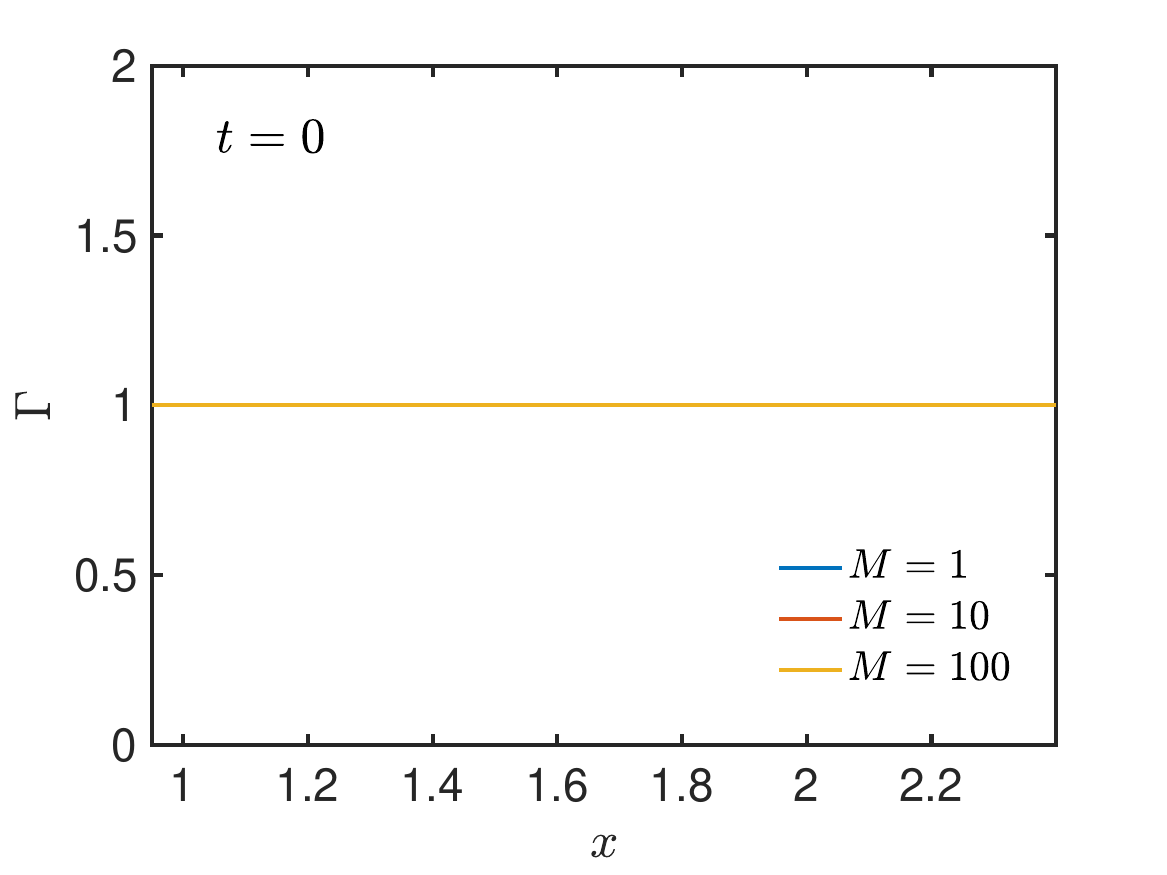}
\caption{}
\label{fig:Gamma vs M t=0}
\end{subfigure}
\begin{subfigure}{0.378\textwidth}
\includegraphics[width=\linewidth]{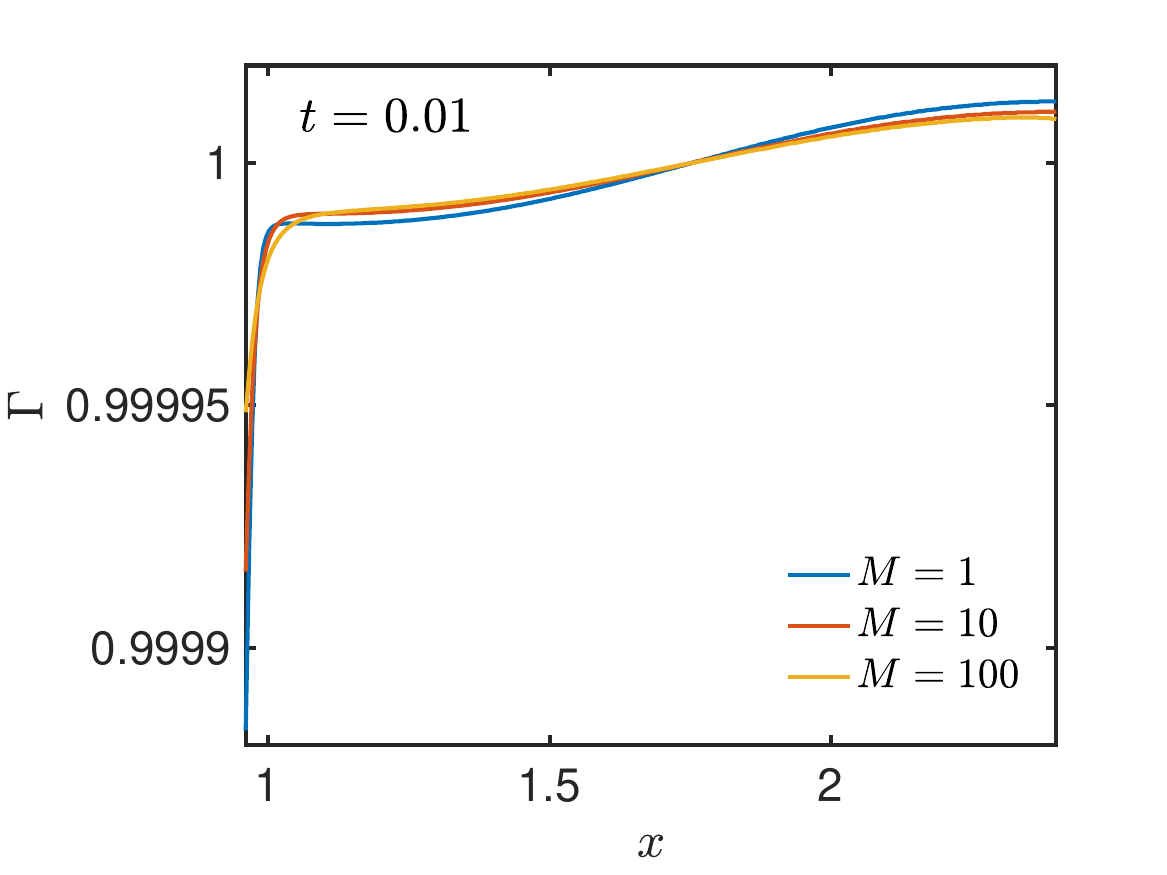}
\caption{}
\label{fig:Gamma vs M t=0.1}
\end{subfigure}
\begin{subfigure}{0.378\textwidth}
\includegraphics[width=\linewidth]{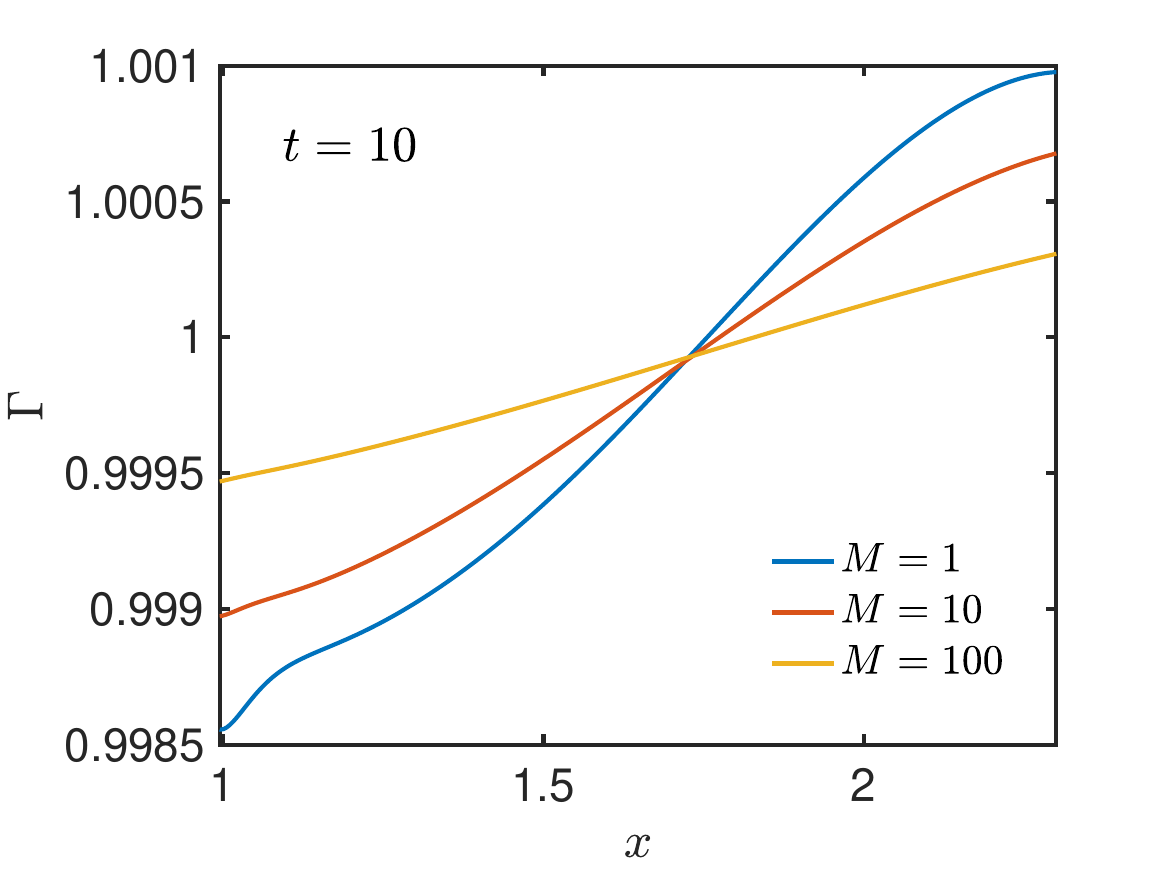}
\caption{}
\label{fig:Gamma vs M t=10}
\end{subfigure}
\begin{subfigure}{0.378\textwidth}
\includegraphics[width=\linewidth]{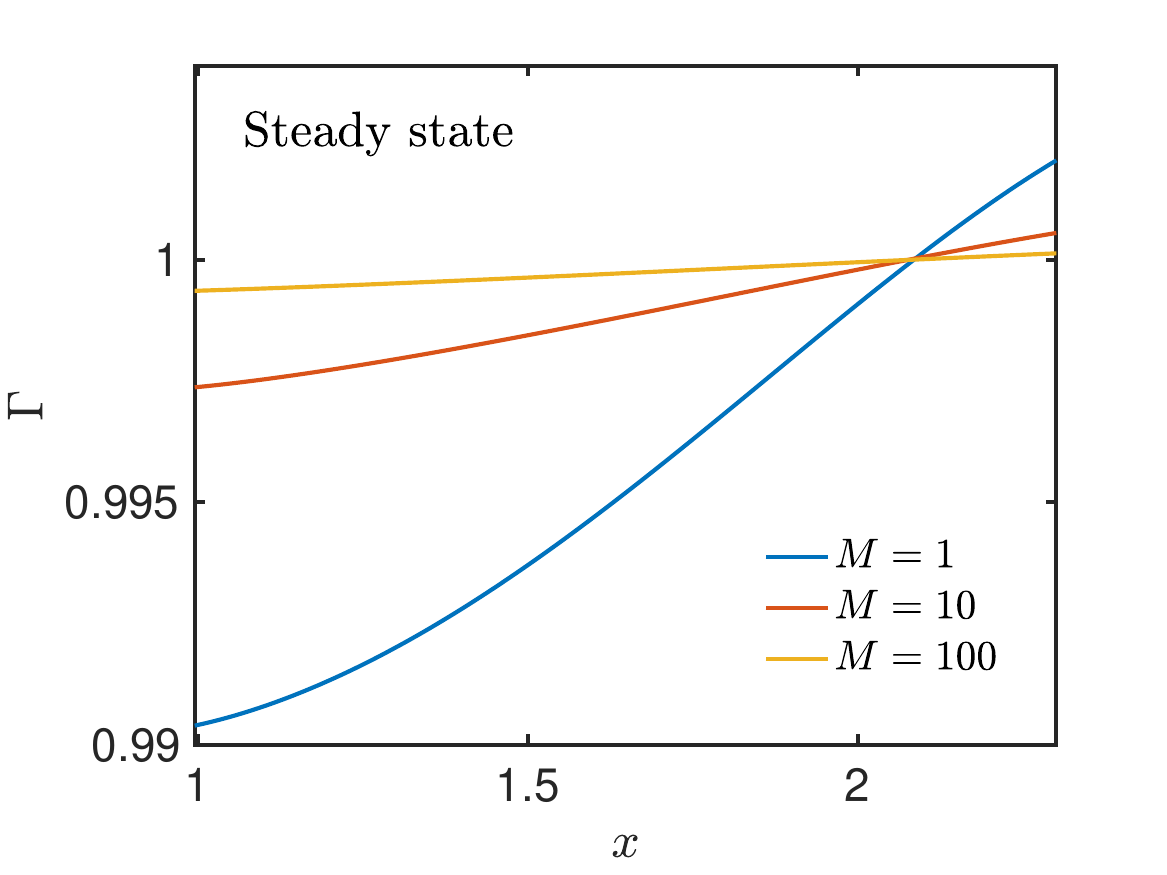}
\caption{}
\label{fig:Gamma vs M t=steady}
\end{subfigure}
\caption{Surfactant-concentration profiles for different $M$ values at (\subref{fig:Gamma vs M t=0})  $t=0$, (\subref{fig:Gamma vs M t=0.1}) $t=0.01$, (\subref{fig:Gamma vs M t=10}) $t=10$, and (\subref{fig:Gamma vs M t=steady}) steady state. The other parameters are $Pe = 1000$, $\Delta P = 0.05$, $L^*=6$, $\mu_r = 0.01$, $v_0 = 0.2$, $\theta_{eq}=10^{\circ}$ ($A=10^5$), and $b=0.001$}
\label{fig:figA1}
\end{figure}

\begin{figure}[ht]
\centering
\begin{subfigure}{0.378\textwidth}
\includegraphics[width=\linewidth]{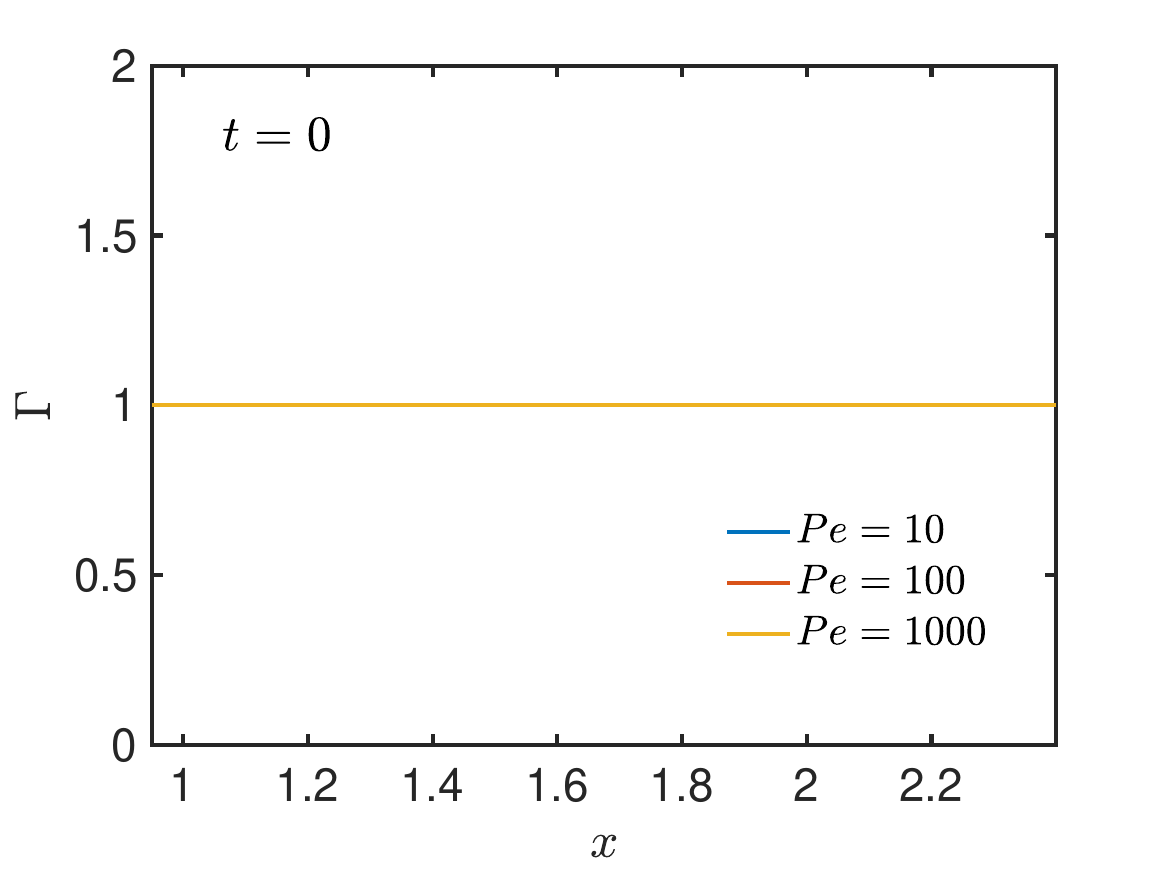}
\caption{}
\label{fig:Gamma vs Pe t=0}
\end{subfigure}
\begin{subfigure}{0.378\textwidth}
\includegraphics[width=\linewidth]{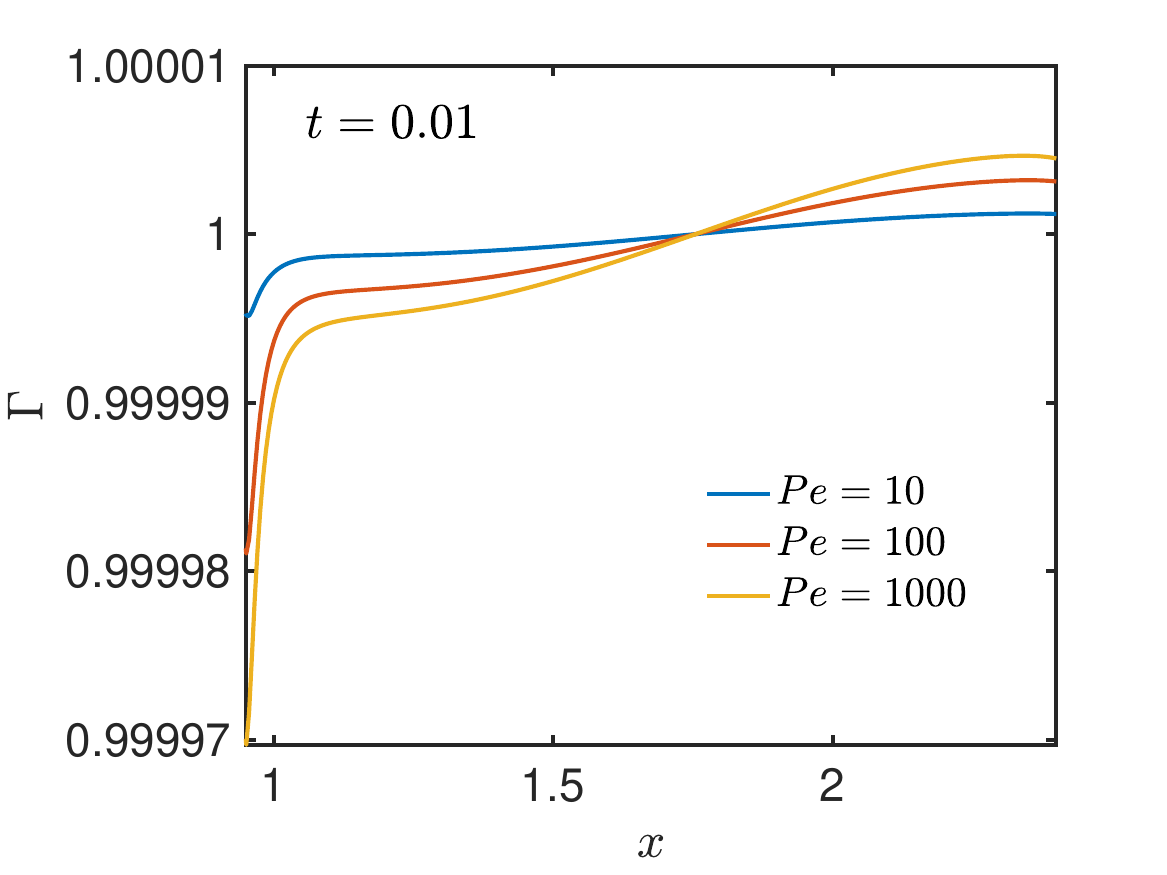}
\caption{}
\label{fig:Gamma vs Pe t=0.1}
\end{subfigure}
\begin{subfigure}{0.378\textwidth}
\includegraphics[width=\linewidth]{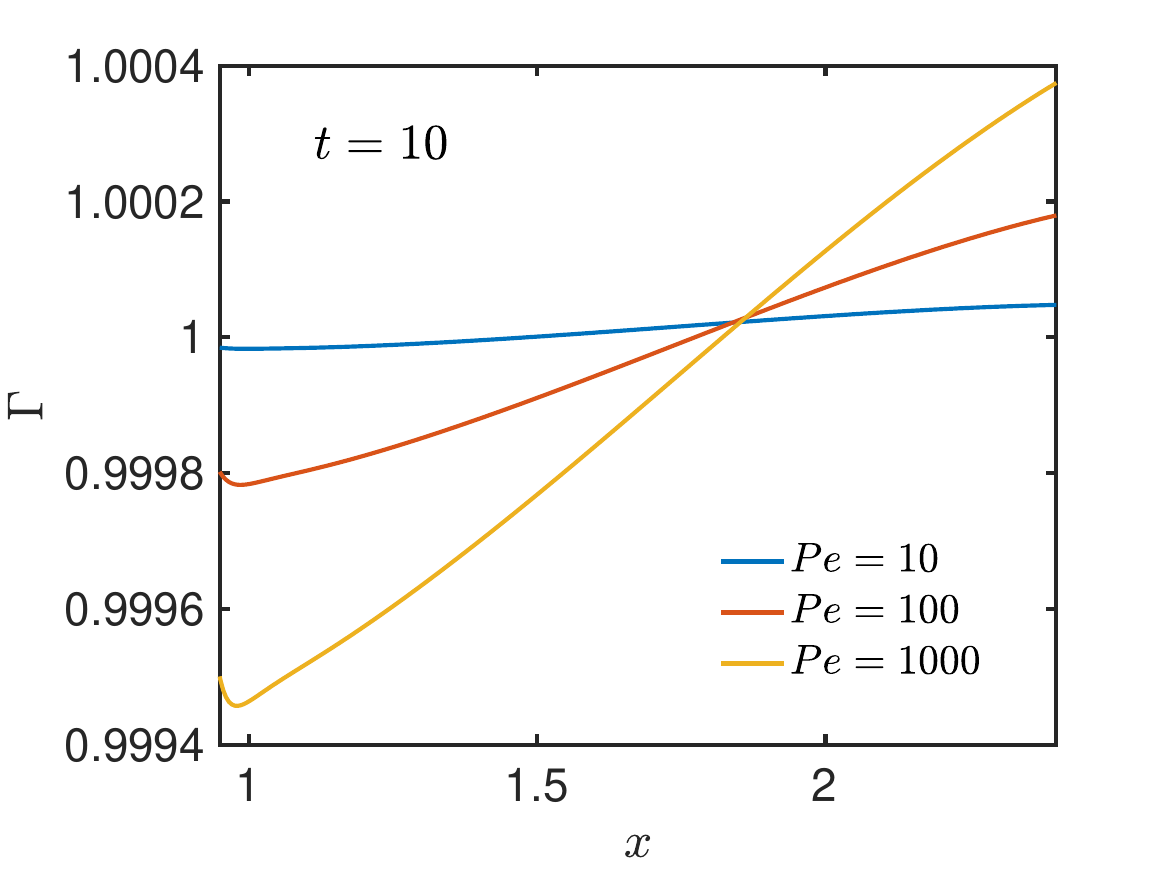}
\caption{}
\label{fig:Gamma vs Pe t=10}
\end{subfigure}
\begin{subfigure}{0.378\textwidth}
\includegraphics[width=\linewidth]{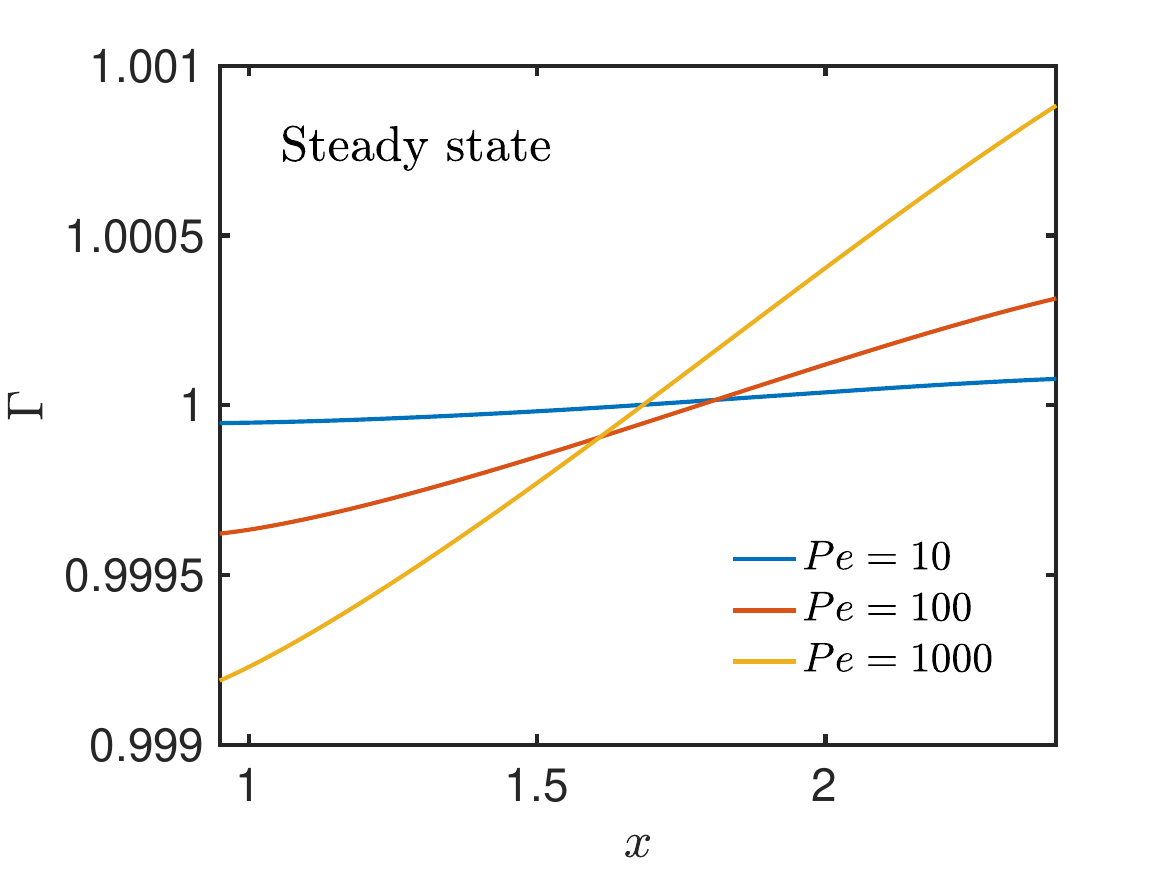}
\caption{}
\label{fig:Gamma vs Pe t=steady}
\end{subfigure}
\caption{Surfactant-concentration profiles for different $Pe$ values at (\subref{fig:Gamma vs Pe t=0})  $t=0$, (\subref{fig:Gamma vs Pe t=0.1}) $t=0.01$, (\subref{fig:Gamma vs Pe t=10}) $t=10$, and (\subref{fig:Gamma vs Pe t=steady}) steady state. The other parameters are $M = 100$, $\Delta P = 0.05$, $L^*=6$, $\mu_r = 0.01$, $v_0 = 0.2$, $\theta_{eq}=10^{\circ}$ ($A=10^5$), and $b=0.001$}
\label{fig:figA2}
\end{figure}

Figure \ref{fig:figA1} shows surfactant concentration profiles for different $M$ values as time progresses, for fixed values of $Pe$ and $\Delta P$. It can be seen that the concentration profiles are almost identical for all $M$ values at $t=0.01$, with the applied pressure gradient leading to accumulation of surfactant near the advancing contact line. This drives Marangoni flow from the advancing to the receding contact line. As $M$ increases, the rate of change of surface tension with respect to the surfactant concentration increases, leading to a stronger Marangoni flow and a flatter concentration profile by  $t=10$ and at steady state.

Figure \ref{fig:figA2} shows surfactant concentration profiles for different $Pe$ values as time progresses, for fixed values of $M$ and $\Delta P$. It can be seen that at $t=0.01$, the smallest surfactant concentration gradient is for $Pe=10$ because surfactant diffusion along the interface becomes important and counters the convection of surfactant toward the advancing contact line. As a result, the concentration profile remains nearly flat at all times for $Pe=10$ and negligible Marangoni flow is generated within the droplet. As $Pe$ increases, convection dominates over diffusion, leading to a larger surfactant concentration gradient and a stronger Marangoni flow from the advancing to the receding contact line. 

\subsection{Influence of reduction in interfacial tension due to presence of surfactant}
\begin{figure}[t]
\centering
\begin{subfigure}{0.49643\textwidth}
\includegraphics[width=\linewidth]{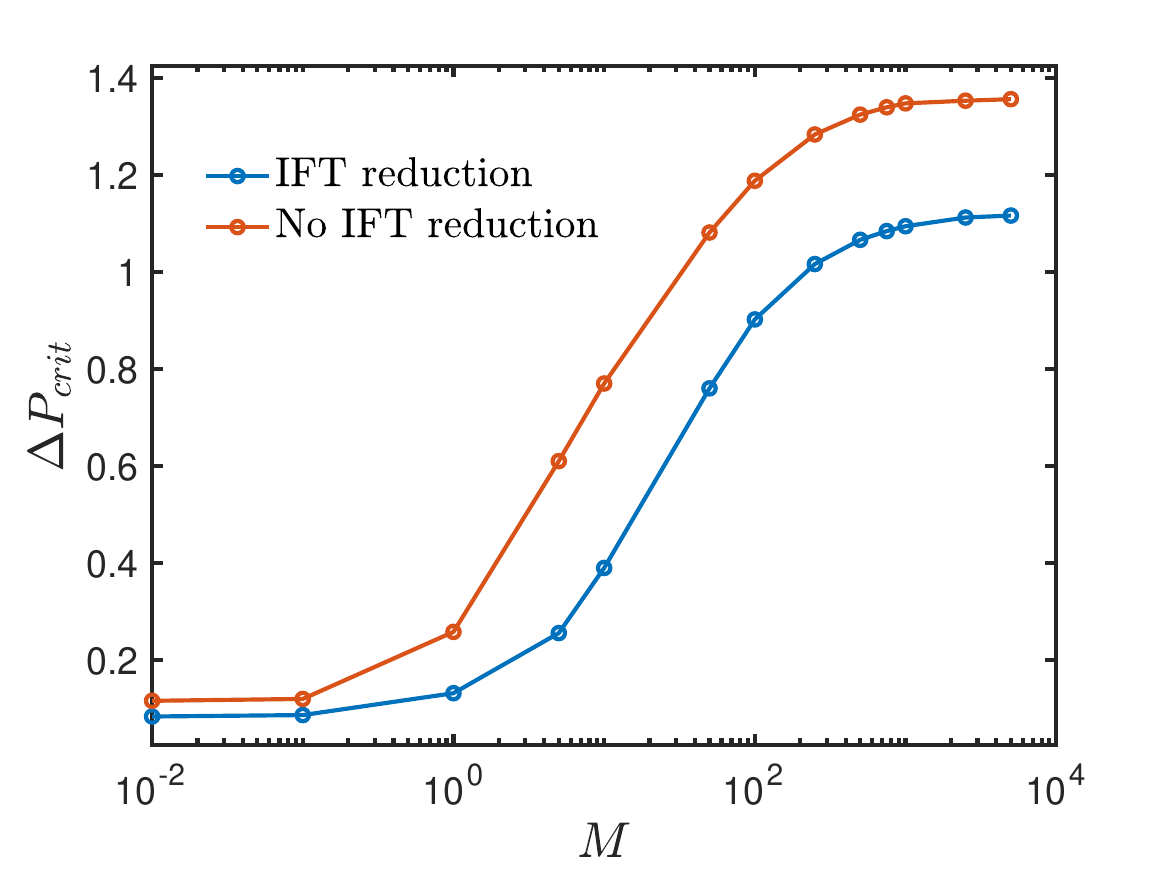}
\caption{}
\label{fig:dP crit vs M IFT vs no IFT reduction}
\end{subfigure}
\hfill
\begin{subfigure}{0.49643\textwidth}
\includegraphics[width=\linewidth]{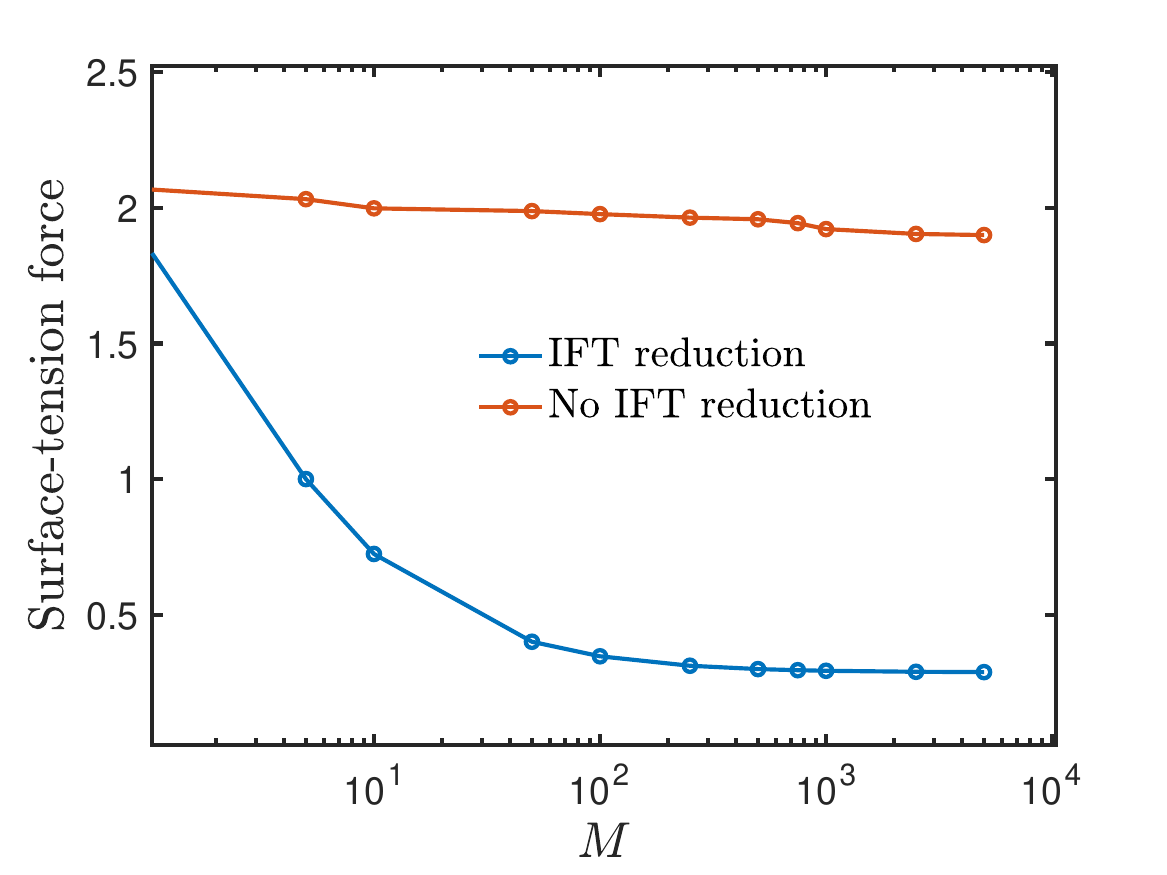}
\caption{}
\label{fig:surf tension force vs M IFT vs no IFT}
\end{subfigure}
\caption{(\subref{fig:dP crit vs M IFT vs no IFT reduction}) $\Delta P_{crit}$ vs. $M$, where the red curve shows the case where there is no change in interfacial tension (IFT) with the surfactant concentration, and the blue curve shows the case which accounts for the change in interfacial tension. (\subref{fig:surf tension force vs M IFT vs no IFT}) Surface-tension force along the droplet contact line vs. $M$ for $\Delta P = 0.05$. The other parameters are $Pe=1000$, $L^*=6$, $\mu_r = 0.01$, $v_0 = 0.2$, $\theta_{eq}=10^{\circ}$ ($A=10^5$), $b=0.001$, $h_d = 0.05h_{max}$, and $w_d=2.5h_d$.}
\label{fig:figA3}
\end{figure}

Figure \ref{fig:dP crit vs M IFT vs no IFT reduction} shows the variation of $\Delta P_{crit}$ with $M$ for $Pe=1000$. The red curve shows results for the case where the reduction in interfacial tension due to the presence of surfactant is neglected by setting $\epsilon^2 \sigma M=0$ in (\ref{eq: normal stress balance}), and the blue curve shows calculations that account for the change in interfacial tension. There is no qualitative difference between the curves, but neglecting the reduction in interfacial tension leads to an overprediction of $\Delta P_{crit}$. This is expected, as a higher interfacial tension leads to a higher surface-tension force (\ref{eq:Dimensionless force balance on smooth substrate}) acting along the contact line for a given $\Delta P$, as shown in figure \ref{fig:surf tension force vs M IFT vs no IFT}, and a larger $\Delta P_{crit}$ is required for droplet depinning.

\end{appendices}

\bibliographystyle{unsrt}
\bibliography{ref}

\end{document}